\documentclass[preprint,prd,showpacs,superscriptaddress,amssymb,amsmath,preprintnumbers,nofootinbib,floatfix]{revtex4-1}
\usepackage[dvipdfmx]{graphicx}
\usepackage{graphicx}
\usepackage{subfigure}
\usepackage{dcolumn}
\usepackage{hyperref}
\usepackage{multirow}
\allowdisplaybreaks[1]

\def\slashchar#1{\setbox0=\hbox{$#1$} 
\dimen0=\wd0 
\setbox1=\hbox{/} \dimen1=\wd1 
\ifdim\dimen0>\dimen1 
\rlap{\hbox to \dimen0{\hfil/\hfil}} 
#1 
\else 
\rlap{\hbox to \dimen1{\hfil$#1$\hfil}} 
/ 
\fi}

\newcommand{\td}{{\rm{d}}}
\newcommand{\e}{{\rm{e}}}
\newcommand{\img}{{\rm{i}}}
\newcommand{\Tr}{{\rm{Tr}}}

\begin{document}
\preprint{
KEK-CP-344,
OU-HET-894
}
\title{
Renormalization of domain-wall bilinear operators with
short-distance current correlators
} 

\author{M.~Tomii}
\email{tomii@post.kek.jp}
\affiliation{
Department of Particle and Nuclear Science,
SOKENDAI (The Graduate University for Advanced Studies), Tsukuba 305-0801, Japan
}
\affiliation{
Theory Center, Institute of Particle and Nuclear Studies,
High Energy Accelerator Research Organization (KEK),
Tsukuba 305-0801, Japan
}

\author{G.~Cossu}
\affiliation{
Theory Center, Institute of Particle and Nuclear Studies,
High Energy Accelerator Research Organization (KEK),
Tsukuba 305-0801, Japan
}

\author{B.~Fahy}
\affiliation{
Theory Center, Institute of Particle and Nuclear Studies,
High Energy Accelerator Research Organization (KEK),
Tsukuba 305-0801, Japan
}

\author{H.~Fukaya}
\affiliation{
Department of Physics, Osaka University,
Toyonaka 560-0043, Japan
}

\author{S.~Hashimoto}
\affiliation{
Department of Particle and Nuclear Science,
SOKENDAI (The Graduate University for Advanced Studies), Tsukuba 305-0801, Japan
}
\affiliation{
Theory Center, Institute of Particle and Nuclear Studies,
High Energy Accelerator Research Organization (KEK),
Tsukuba 305-0801, Japan
}

\author{T.~Kaneko}
\affiliation{
Department of Particle and Nuclear Science,
SOKENDAI (The Graduate University for Advanced Studies), Tsukuba 305-0801, Japan
}
\affiliation{
Theory Center, Institute of Particle and Nuclear Studies,
High Energy Accelerator Research Organization (KEK),
Tsukuba 305-0801, Japan
}

\author{J.~Noaki}
\affiliation{
Theory Center, Institute of Particle and Nuclear Studies,
High Energy Accelerator Research Organization (KEK),
Tsukuba 305-0801, Japan
}

\collaboration{JLQCD collaboration}
\noaffiliation

\begin{abstract}
We determine the renormalization constants for flavor non-singlet fermion
bilinear operators of M\"obius domain-wall fermions.
The renormalization condition is imposed on the correlation functions in the
coordinate space, such that the non-perturbative lattice calculation reproduces
the perturbatively calculated counterpart at short distances.
The perturbative expansion is precise as the coefficients are available up to
$O(\alpha_s^4)$.
We employ $2+1$-flavor lattice ensembles at three lattice spacings in the
range 0.044--0.080~fm.
\end{abstract}

\maketitle

\section{Introduction}
\label{sec:intro}

Renormalization of lattice operators is a necessary step for the lattice QCD
calculations of scheme-dependent quantities such as the matrix elements of the
weak effective Hamiltonian, which contains composite operators of
quark fields such as bilinear and four-fermion operators.
In phenomenological studies, comparison of lattice results with experimental
values or theoretical results from other approaches has to be performed with a
matched renormalization scheme and scale.
The conventional choice for this is the $\rm\overline{MS}$ scheme.
Among bilinear operators, vector (and axial-vector) currents are simplest because
they are not renormalized because of current conservation.
On the lattice, however, the commonly used local vector current is not conserved
at finite lattice spacing and a finite renormalization to match the continuum
counterpart is necessary.
The axial-current is renormalized differently unless the lattice fermion formulation
respects chiral symmetry.
Scale dependent operators such as scalar and pseudoscalar currents require
renormalization in any case.

Matching to the continuum renormalization scheme necessarily involves
perturbation theory, since the $\rm\overline{MS}$ scheme adopted in
the continuum theory is inherently perturbative.
It means that the renormalization condition must be applied at momentum
scale of 2~GeV or higher, where QCD becomes perturbative.
In this regime, the discretization effect of lattice theory becomes significant as
typical value of lattice cutoff for currently available ensembles is 2--3~GeV.
This causes the so-called window problem, which implies that one has to identify
a window of momentum scale where the systematic uncertainties of perturbative
and lattice calculations are controlled.
One of the widely used method is the non-perturbative renormalization through the
RI/MOM scheme as an intermediate renormalization condition
\cite{Martinelli:1994ty}.
In this method, one calculates the vertex function involving the composite operator
of interest with a certain momentum configuration of external quarks in a fixed gauge.
The external momenta are carefully chosen to satisfy the window condition as mentioned
above.
In many cases, one can find the region where the discretization effect is under control
by inspecting the momentum dependence of the obtained vertex function.
The convergence of the perturbative expansion, on the other hand, is not well tested,
as the perturbative expansion is available only at the one- or two-loop level in many cases.
In fact, this is a dominant source of systematic uncertainty in this method.

In this work, we adopt another renormalization condition, which is called the
coordinate space method or the X-space method.
The method was first proposed in \cite{Martinelli:1997zc} 
and has so far been applied in \cite{Gimenez:2004me,Cichy:2012is}.
In this method, the renormalization condition is imposed on the correlation function
at a certain distance in the coordinate space.
The correlation functions are those of composite operators of interest.
A typical example is a correlation function of vector currents $\bar u\gamma_\mu d(x)$
and $\bar d\gamma_\nu u(y)$ placed at certain points $x$ and $y$ in the coordinate
space.
Here, $u$ and $d$ denote the up and down quark fields, respectively.
The correlation function thus constructed
$\langle\bar u\gamma_\mu d(x) \bar d\gamma_\nu u(y)\rangle$ is gauge invariant,
and more importantly it is a common object in the continuum QCD for which the
state-of-the-art techniques of perturbative QCD have been applied.
For the (axial-)vector and (pseudo)scalar correlators, the perturbative series has
been calculated to the order of $\alpha_s^4$ \cite{Chetyrkin:2010dx}, with which
the uncertainty due to a truncation of the perturbative series is highly suppressed and
the remaining uncertainty is reliably estimated using the convergence behavior of the
series.
Another important advantage of the X-space method is that the X-space correlator
does not involve extra divergences other than those of the operators involved, because
of the finite distance between $x$ and $y$. The corresponding correlator in the
momentum space has an extra divergence from a loop contracting the two points,
and one has to consider a derivative in the momentum space to extract a well-defined
quantity.

The window problem does exist for the X-space method.
The distance between $x$ and $y$ must be sufficiently small for perturbation
theory to be convergent, and at the same time it has to be large compared to
the lattice spacing $a$.
The main point of this work is therefore to identify such a region.
On the perturbative side, we investigate the convergence property of the
perturbative series known to $\alpha_s^4$.
The convergence can be improved by optimizing the scale $\mu$ of coupling
constant $\alpha_s(\mu)$, depending on the distance $|x-y|$.
We first attempt the use of the Brodsky-Lepage-Mackenzie (BLM)
scale setting procedure \cite{Brodsky:1982gc},
which in fact leads to a much improved convergence compared to a na\"\i ve
choice of $\mu = 1/|x-y|$.
In order to estimate the remaining uncertainty due to the truncation of higher order
terms, we vary the scale in some range and inspect the stability of the result.
A detailed account of this study is given in Section~\ref{sec:cont_th} for both
vector and scalar correlators.

As we include the longer-distance correlators in the analysis,
non-perturbative power corrections could also appear as well as the
problem of increasingly less convergent perturbative series.
Such effect is described by the Operator Product Expansion (OPE)
\cite{Shifman:1978bx}.
The leading contribution comes from the dimension-four operators: the
quark condensate and gluon condensate.
The quark condensate appears with a quark mass, and induces a linear
dependence on the quark mass.
We confirm that the lattice data are consistent with
the prediction of OPE.
We can then eliminate this type of power correction by combining vector
and axial-vector correlators with appropriate factors.
Similar method is also applied for scalar and pseudoscalar correlators.
The effect of gluon condensate is more difficult to identify.
We fit the lattice data to the corresponding functional form in $|x-y|$.

At short distances on the lattice, the discretization error is a major concern,
which we investigate in detail in Section~\ref{sec:lat_calc}.
The discretization effect is indeed very substantial at the distance of
$|x-y|/a \lesssim$ 10.
Fortunately, most of such effects can be eliminated by subtracting the correlators
calculated at the tree-level (or mean-field improved).
Further reduction can be achieved by selecting the direction of the points on the lattice.
These improvements have already been applied in the previous studies
\cite{Gimenez:2004me,Cichy:2012is}, and we refine them to minimize the
remaining errors.
Even after such reductions of errors, the discretization effects are still significant
at the level of 3--5\% for the correlators.
We attempt to eliminate them by using the lattice data available for us at three
different lattice spacings in the range 0.044--0.080~fm.
We fit the lattice data at short distances assuming that the leading discretization
effect is $O(a^2)$, and subtract it.
The remaining error is then at the level of one per cent.

In this work, we determine the renormalization constants for the vector
(and axial-vector) current and scalar (and pseudoscalar) density operator
composed of the M\"obius domain-wall fermions
\cite{Brower:2004xi,Brower:2012vk}.
M\"obius domain-wall fermion is an improved implementation of the
domain-wall fermion \cite{Kaplan:1992bt,Shamir:1993zy}.
The four-dimensional effective Dirac operator of the
domain-wall fermion precisely satisfies the Ginsparg-Wilson relation and thus
respects the modified chiral symmetry on the lattice.
It implies that the vector and axial-vector renormalization factors are equal to each
other to a good precision. With the M\"obius implementation, the residual mass, which
quantifies the size of the Ginsparg-Wilson violation, is at the order of 1~MeV
or less on our lattice ensembles. For the determination of the renormalization
constants, this amount of violation can be safely neglected.
In fact, we confirm that the corresponding current correlators agree very
precisely at short distances where the effect of the quark condensate is negligible.

This paper is organized as follows.
In Section~\ref{sec:second}, we describe the basic strategy to determine the
renormalization constants.
In Section~\ref{sec:cont_th}, we discuss the detail of massless
correlators in perturbation theory and the improvement of their convergence.
After that, we give the contribution of OPE to the correlators.
In Section~\ref{sec:lat_calc}, we show the analyses of lattice calculation
including our lattice setup and some managements of lattice artifacts.
In Section~\ref{sec:NPR}, we explain the detail of determination of the
renormalization factors and show the final results.

\section{Renormalization condition in the X-space method}
\label{sec:second}

In the X-space method, the renormalization constants are determined by
analyzing two-point correlation functions at finite separation $x-y$.
In this work, we consider the following four channels of flavor non-singlet
correlators in the coordinate space,
\begin{equation}
\begin{array}{ll}
\Pi_S(x) = \langle S(x)S(0)^\dag\rangle,
& \hspace{10mm}
\Pi_P(x) = \langle P(x)P(0)^\dag\rangle,
\\
\Pi_V (x) = \sum_\mu \langle V_\mu(x)V_\mu(0)^\dag\rangle,
& \hspace{10mm}
\Pi_A (x) = \sum_\mu \langle A_\mu(x)A_\mu(0)^\dag\rangle,
\end{array}
\label{eq:def_correl}
\end{equation}
where each operator is composed of (mass degenerate) up and down quark
fields $u(x)$ and $d(x)$,
\begin{equation}
\begin{array}{ll}
S(x) = \bar u(x)d(x),
& \hspace{10mm}
P(x) = \bar u(x)\img\gamma_5d(x),
\\
V_\mu(x) = \bar u(x)\gamma_\mu d(x),
& \hspace{10mm}
A_\mu(x) = \bar u(x)\gamma_\mu\gamma_5d(x).
\end{array}
\label{eq:def_currents}
\end{equation}
Here, the coordinate $y$ of the source point is fixed at origin
and correlators are parametrized by a four-dimensional coordinate $x$
with an assumption of translational invariance.

We renormalize the quark bilinear operators on the lattice to
those in the $\rm\overline{MS}$ scheme at a renormalization scale,
which is often set to 2 or 3~GeV.
Neglecting the contributions of irrelevant operators, the renormalization
is multiplicative, {\it i.e.}
\begin{equation}
O_\Gamma^{\rm\overline{MS}}(2{\rm\ GeV};x)
= Z_\Gamma^{\rm\overline{MS}/lat}(2{\rm\ GeV},a)
O_\Gamma^{\rm lat}(a;x),
\end{equation}
where $\Gamma \in \{S,P,V,A\}$, $O_\Gamma \in \{S, P, V_\mu, A_\mu\}$,
and $Z_\Gamma^{\rm\overline{MS}/lat}(2{\rm\ GeV},a)$ is the renormalization constant.
The renormalization condition in the X-space method is imposed by requiring
\begin{equation}
\Big(\widetilde Z_\Gamma^{\rm\overline{MS}/lat}(2{\rm\ GeV},a;x)\Big)^2
\Pi_\Gamma^{\rm lat}(a;x) = \Pi_\Gamma^{\rm\overline{MS}}(2{\rm\ GeV};x),
\end{equation}
or
\begin{equation}
\widetilde Z_\Gamma^{\rm\overline{MS}/lat}(2{\rm\ GeV},a;x)
= \sqrt{\Pi_\Gamma^{\rm\overline{MS}}(2{\rm\ GeV};x)\over\Pi_\Gamma^{\rm lat}(a;x)},
\label{eq:zfactor_cond}
\end{equation}
at a finite distance $x$.
Note that $\widetilde Z_\Gamma^{\rm\overline{MS}/lat}(2{\rm\ GeV},a;x)$
still contains some dependence on $x$.
It originates from errors arising in
the continuum $\Pi_\Gamma^{\rm\overline{MS}}(2{\rm\ GeV};x)$ and lattice
$\Pi_\Gamma^{\rm lat}(a;x)$ correlators.
The continuum one suffers from truncation of the perturbative expansion
as we discuss in the following sections.
On the other hand, the lattice correlator contains discretization effects.
In addition, the $x$-dependence of
$\widetilde Z_\Gamma^{\rm\overline{MS}/lat}(2{\rm\ GeV},a;x)$ is
also caused by non-perturbative effects at large $|x|$ in full QCD,
which are not encoded in the continuum perturbative correlator.
In order to extract the renormalization constant, which must be independent of $x$,
the distance $|x|$ of correlators should be chosen in a window
$a\ll |x|\ll 1/\Lambda_{\rm QCD}$ to suppress these possible errors.
In Section~\ref{sec:NPR}, the systematic effects arising in
$\widetilde Z_\Gamma^{\rm\overline{MS}}({2\rm\ GeV},a;x)$ are discussed
in more detail and the renormalization factor
$Z_\Gamma^{\rm\overline{MS}}({2\rm\ GeV},a)$ is determined.

\section{Continuum theory}
\label{sec:cont_th}

\subsection{Massless perturbation theory}
\label{subsec:massless_pt}

In this subsection, we discuss the convergence of the perturbative
expansion of the massless correlators.
Since the scalar and pseudoscalar correlators, as well as the
vector and axial-vector correlators, degenerate
in the massless perturbation theory, {\it i.e.}
$\Pi_S = \Pi_P$ and $\Pi_V=\Pi_A$,
we consider only the two channels $\Pi_S$ and $\Pi_V$.

The perturbative expansion of the vector correlator is written as
\begin{equation}
\Pi_V^{\rm\overline{MS}}(x)
= {6\over\pi^4x^6}
\left(1+\sum_{i=1}^\infty C_i^V a_s(\mu_x)^i
\right),
\end{equation}
with perturbative coefficients $C_i^V$.
Here, $a_s(\mu_x) = \alpha_s(\mu_x)/\pi$ is the strong coupling constant and
its scale $\mu_x$ is set as
\begin{equation}
\mu_x = {1\over |x|}.
\end{equation}

One can reorganize the perturbative series using the renormalization group
to another scale $\mu_x^*$, {\it i.e.}
\begin{equation}
\Pi_V^{\rm\overline{MS}}(x)
= {6\over\pi^4x^6}\left(
1+\sum_{i=1}^\infty C_i^V(\mu_x^*)a_s(\mu_x^*)^i
\right),
\end{equation}
which is exact when the perturbative series includes all orders,
provided that the coefficients $C^V_i(\mu^*_x)$ are converted appropriately.
The conversion formula of the coupling constants is available to four-loop level
\cite{vanRitbergen:1997va}.

Chetyrkin and Maier \cite{Chetyrkin:2010dx} wrote the perturbative
coefficients up to $O(a_s^4)$ at the renormalization scale
$\mu_x$ in the $\rm\widetilde{MS}$ scheme,
which is the same as the $\rm\overline{MS}$ scheme at a scale
$\tilde \mu_x = 2\e^{-\gamma_E}\mu_x \simeq 1.123/|x|$.
Here, $\gamma_E = 0.5772\ldots$ is Euler's constant.
In our notation, these coefficients correspond to
$C_i^V(\tilde\mu_x)$.
The perturbative coefficients
$C_i^V(\mu_x^*)$ at $\mu_x^* = \mu_x, \tilde\mu_x$ obtained by using the
results of \cite{Chetyrkin:2010dx} and \cite{vanRitbergen:1997va}
are summarized in Appendix~\ref{sec:app_pert}.

Since the perturbative expansion truncated at a finite order may depend
on $\mu_x^*$, there is an optimal choice for the scale that leads to a good
convergence.
One possible recipe to choose the optimal scale $\mu_x^*$ is the
Brodsky-Lepage-Mackenzie (BLM) approach \cite{Brodsky:1982gc},
which is motivated by an idea of absorbing the
higher order contributions of gluon vacuum polarization into the coupling
constant.
The scale is chosen such that the perturbative coefficient at $a_s^2$
becomes independent of the number of flavors $n_f$.
This BLM scale $\mu_x^{\rm BLM}$ of the vector correlator thus determined is
\begin{equation}
\mu_x^{\rm BLM} =
2\exp\left[{1\over2}(4\zeta_3-3-2\gamma_E)\right]\mu_x
\simeq {2.7733\over |x|},
\label{eq:naive_BLM}
\end{equation}
where $\zeta_3 \simeq 1.2021$.
The perturbative coefficients
$C_i^V(\mu_x^{\rm BLM})$ at the BLM scale
are summarized in Appendix~\ref{sec:app_pert}.

\begin{figure}[t]
\begin{center}
\includegraphics[width=120mm]{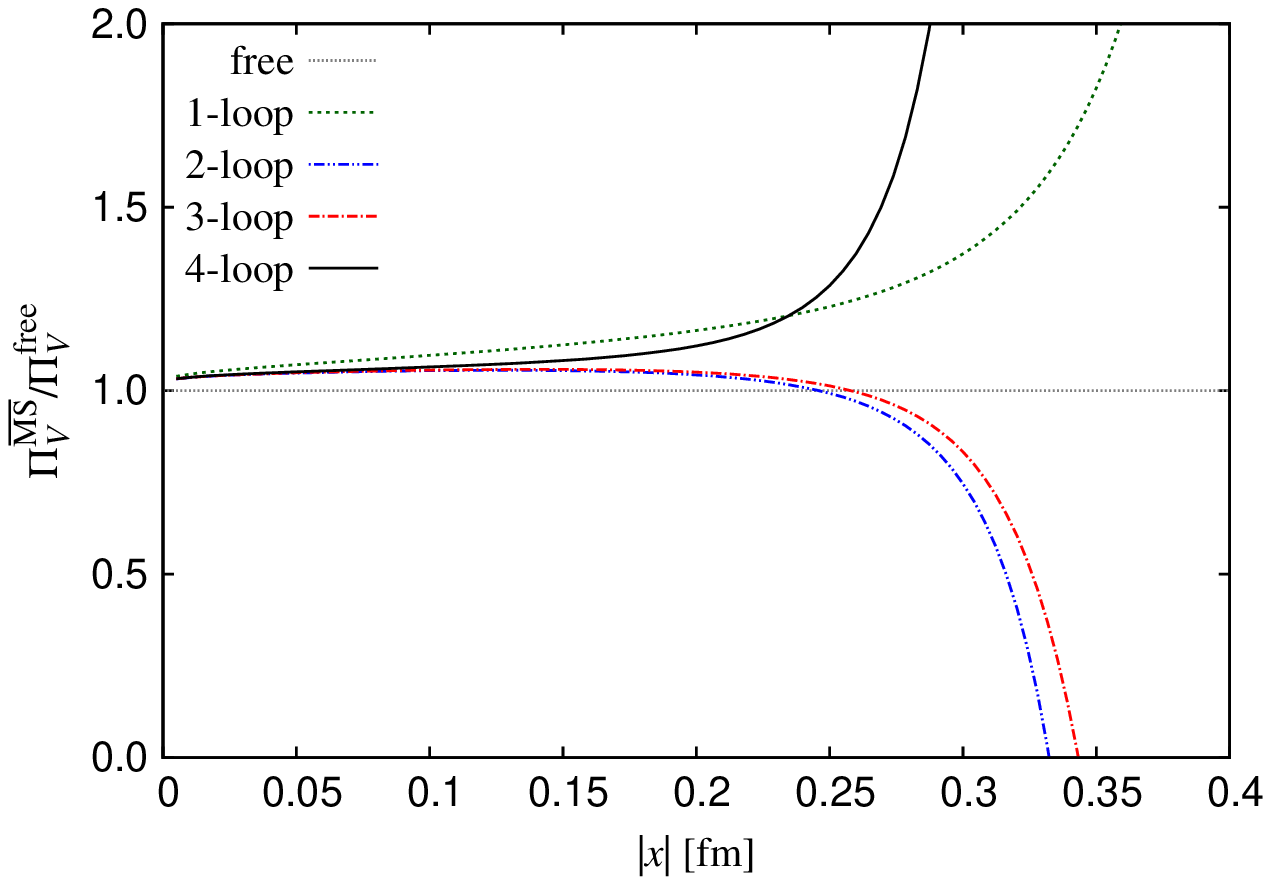}
\caption{
Perturbative expansion of the vector correlator renormalized in
the $\rm\overline{MS}$ scheme with $n_f = 3$.
The results at $\mu_x^* = \mu_x$
truncated at $a_s^0$ (fine-dotted),
$a_s$ (dotted), $a_s^2$ (dashed double-dotted), $a_s^3$ (dashed dotted),
and $a_s^4$ (solid) are plotted as functions of $|x|$.
}
\label{fig:vv_l0.0}
\end{center}
\begin{center}
\includegraphics[width=120mm]{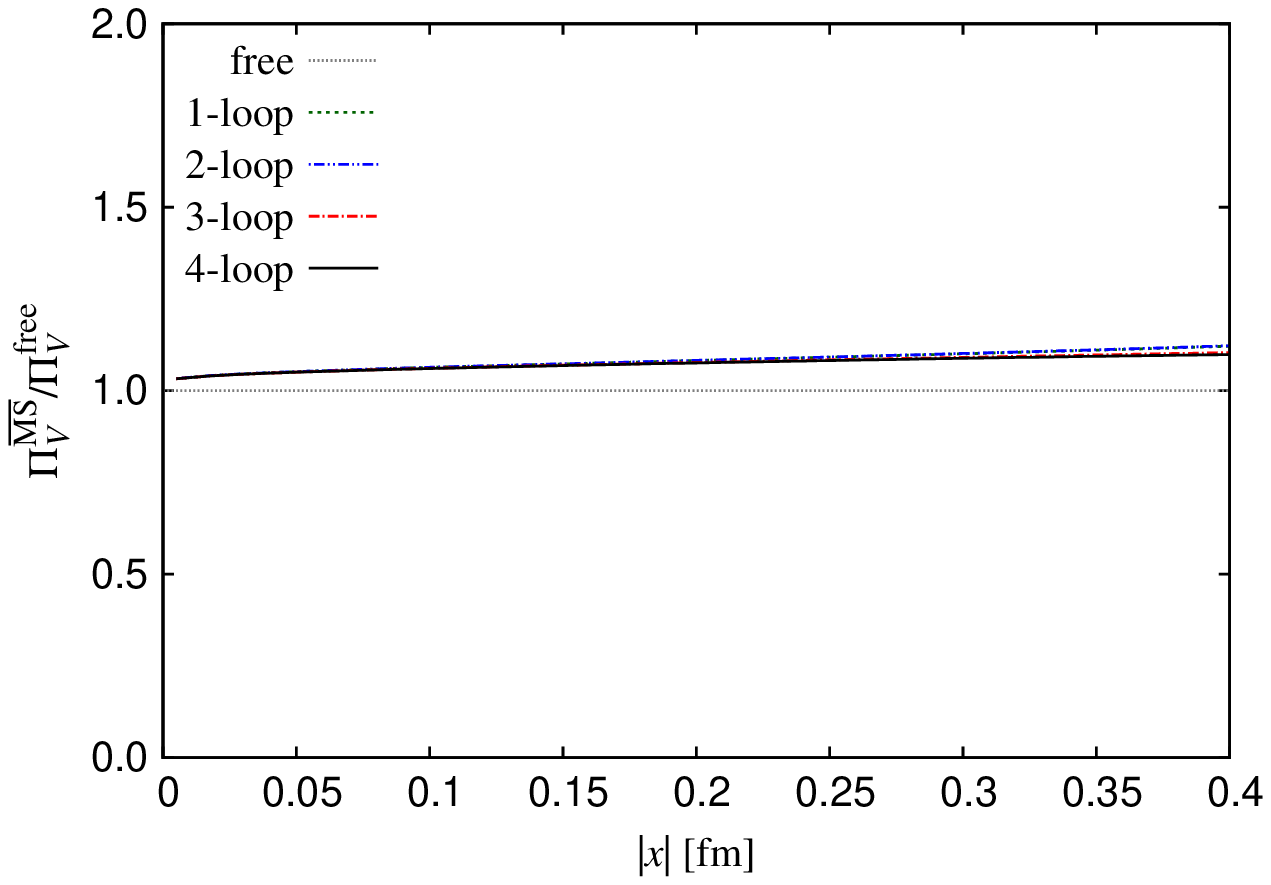}
\caption{
Perturbative expansion of the vector correlator renormalized in
the $\rm\overline{MS}$ scheme with $n_f = 3$.
The results at $\mu_x^* = \mu_x^{\rm BLM}$
truncated at $a_s^0$ (fine-dotted),
$a_s$ (dotted), $a_s^2$ (dashed double-dotted), $a_s^3$ (dashed dotted),
and $a_s^4$ (solid) are plotted as functions of $|x|$.
}
\label{fig:vv_BLM}
\end{center}
\end{figure}

Figure~\ref{fig:vv_l0.0} shows the vector correlator
calculated with $\mu_x^* = \mu_x$ and $n_f=3$.
It is normalized by the tree-level correlator $\Pi_V^{\rm free}(x) = 6/\pi^4x^6$.
A reasonable convergence is observed only below
$|x| \sim 0.15$~fm, which is not so large compared to our lattice spacings
$a = $ 0.044--0.080~fm.
It implies that there is no renormalization window satisfying the condition that
the perturbative calculation is convergent and the discretization effects are
sufficiently small.
The convergence of the perturbative series at $\mu_x^* = \tilde\mu_x$ is
similar to that at $\mu_x^* = \mu_x$.
On the other hand, the result at $\mu_x^* = \mu_x^{\rm BLM}$
with $n_f=3$ shows much better convergence
as plotted in Fig.~\ref{fig:vv_BLM},
implying that the convergence of the perturbative series is actually
improved by tuning the renormalization scale $\mu_x^*$.

\begin{figure}[t]
\begin{center}
\subfigure{\mbox{\raisebox{1mm}{\includegraphics[width=81mm]{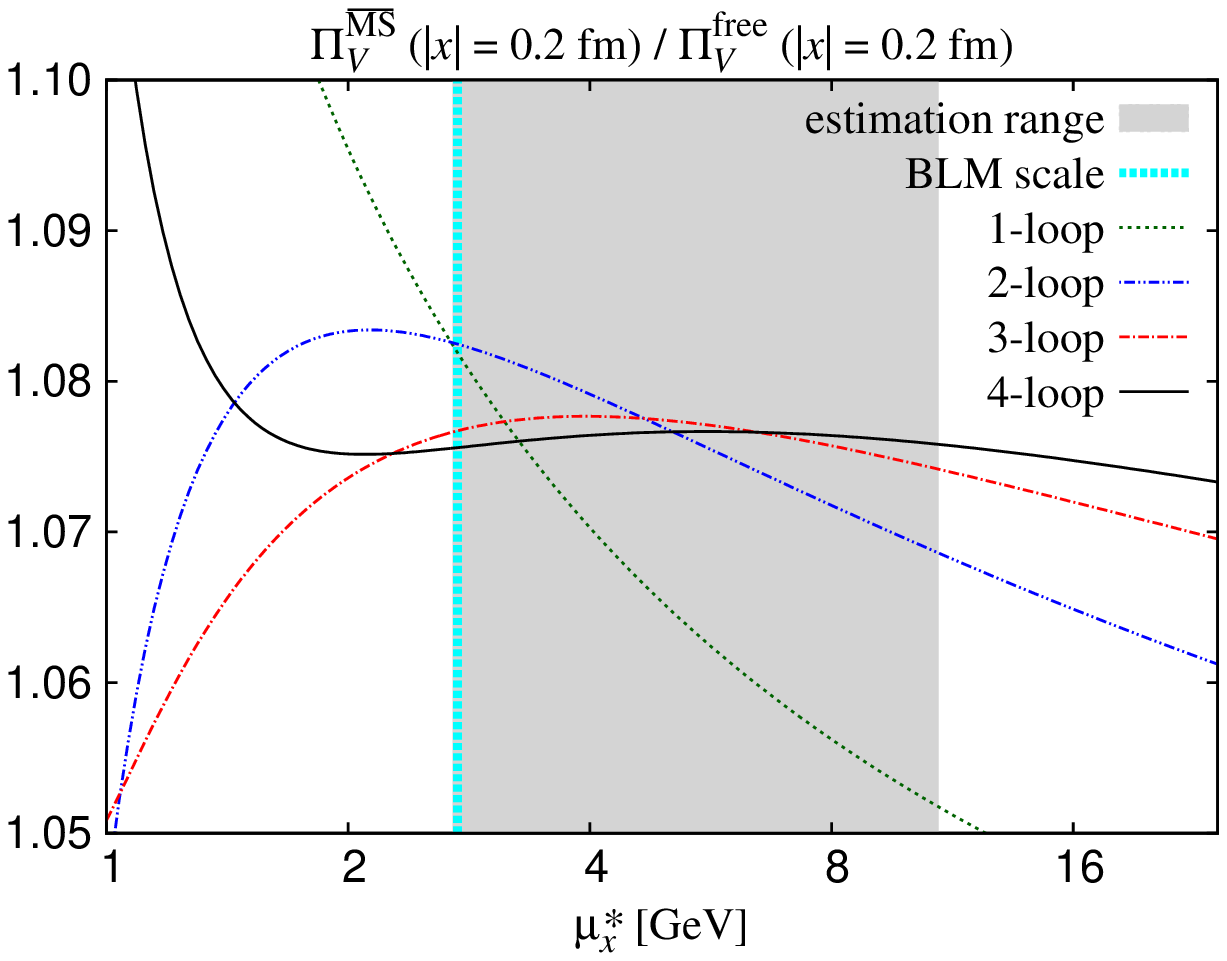}}}}
\subfigure{\mbox{\raisebox{1mm}{\includegraphics[width=81mm]{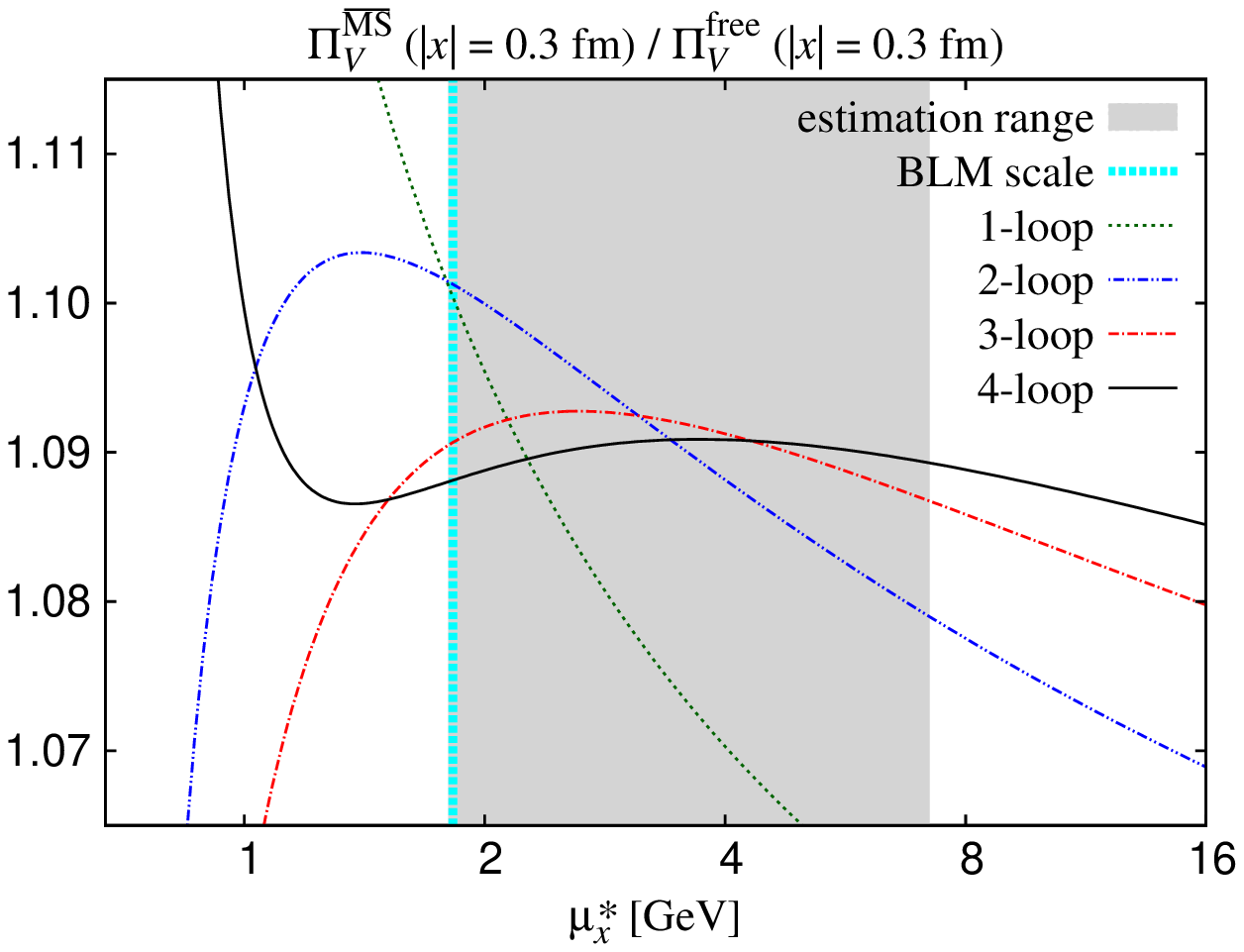}}}}
\subfigure{\mbox{\raisebox{1mm}{\includegraphics[width=81mm]{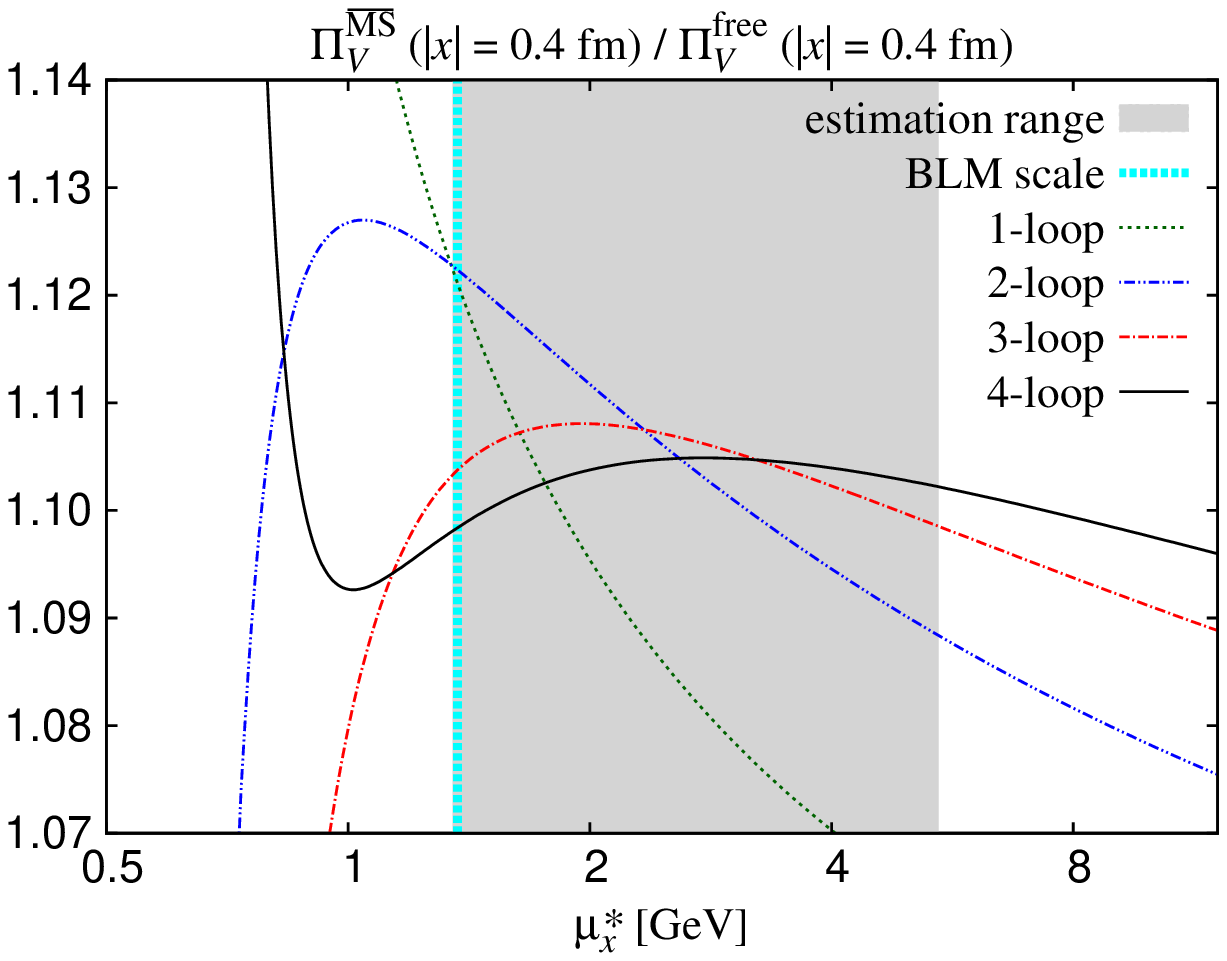}}}}
\subfigure{\mbox{\raisebox{1mm}{\includegraphics[width=81mm]{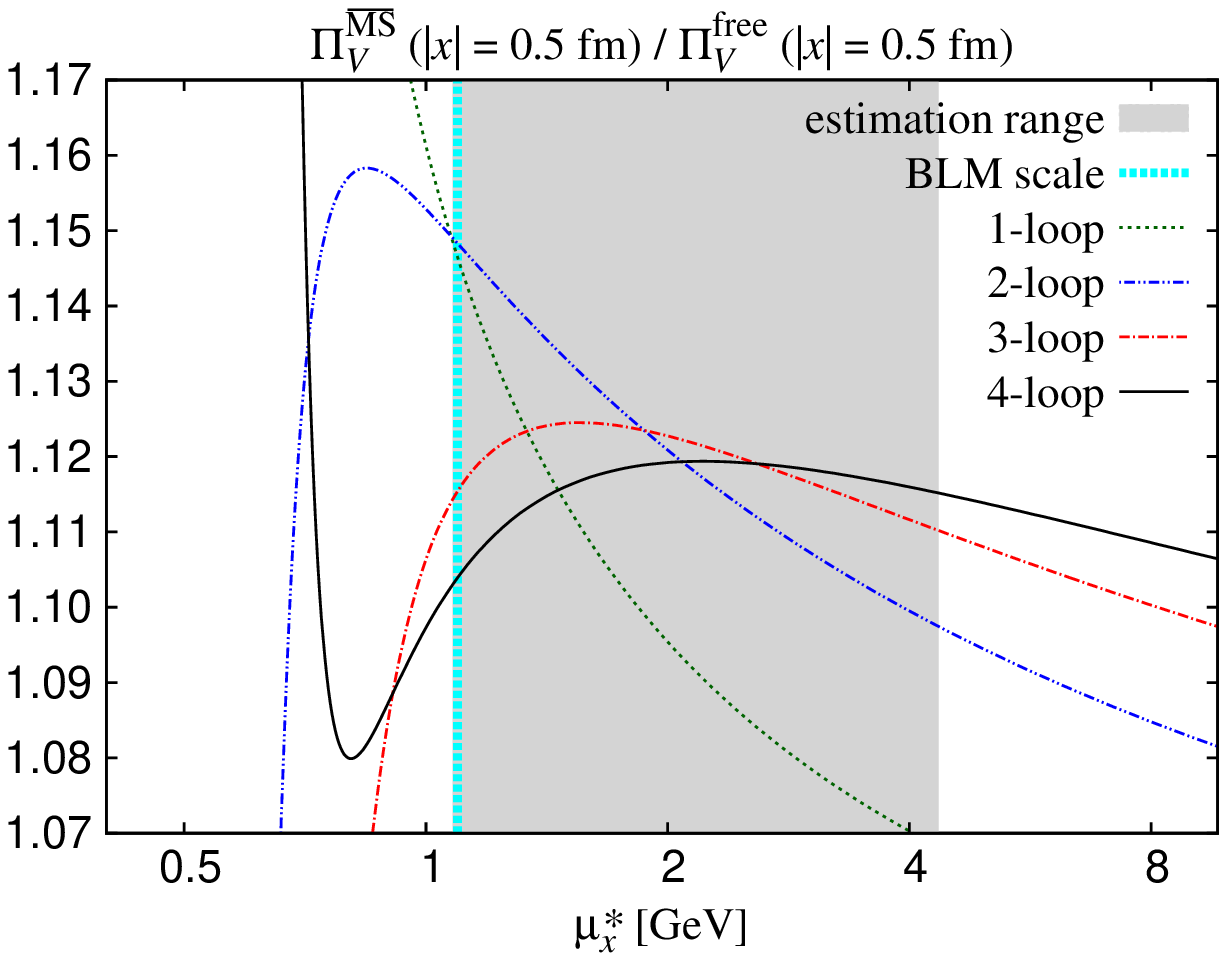}}}}
\caption{
Vector correlator renormalized in the $\rm\overline{MS}$ scheme
at the specific distances 0.2~fm (top/left), 0.3~fm (top/right),
0.4~fm (bottom/left), and 0.5~fm (bottom/right) as functions
of $\mu^*_x$.
The results with $n_f=3$ truncated at
$a_s$ (dotted), $a_s^2$ (dashed double-dotted), $a_s^3$ (dashed dotted),
and $a_s^4$ (solid) are plotted.
The gray band represents the region in which we estimate the uncertainty
of the higher order corrections.
The vertical bold line near the lower end of the gray band stands for
the BLM scale (\ref{eq:naive_BLM}).
}
\label{fig:ldps_vv}
\end{center}
\end{figure}

In order to choose the optimal scale and estimate the
uncertainty of the higher order corrections, we investigate the
$\mu_x^*$-dependence of the perturbative calculation.
The $\mu_x^*$-dependence of the vector correlator with $n_f=3$ is shown
in Fig.~\ref{fig:ldps_vv} for several distances in the range 0.2--0.5~fm.
Since the all-order calculation has to be independent of $\mu_x^*$,
we determine the optimal scale $\mu_x^{\rm *,opt}$ as the value which
minimizes the $\mu_x^*$-derivative of the four-loop correlator,
\begin{equation}
\mu_x^{\rm *,opt} = \e^{1.7}\mu_x \simeq {5.5\over |x|}.
\label{eq:vv_best_l}
\end{equation}
The uncertainty of the higher order corrections to the vector correlator
is estimated by varying $\mu_x^*$ in the region
$[{1\over2}\mu_x^{\rm *,opt},2\mu_x^{\rm *,opt}]$,
which is shown in Fig.~\ref{fig:ldps_vv} by the gray band.

\begin{figure}[t]
\begin{center}
\includegraphics[width=120mm]{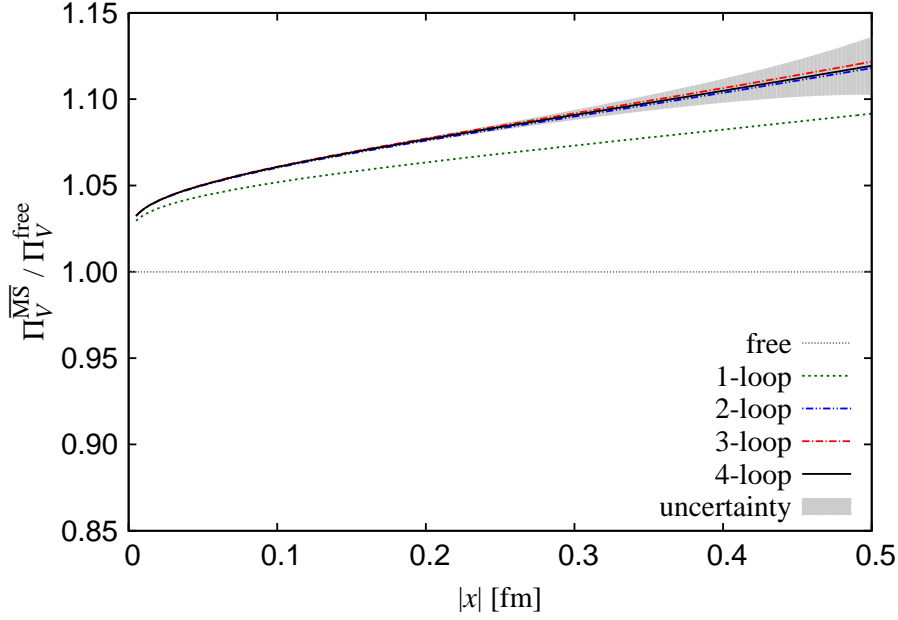}
\caption{
Perturbative expansion of the vector correlator renormalized in
the $\rm\overline{MS}$ scheme with $n_f = 3$.
The results of  the perturbative series (\ref{eq:vv_best_l_nf3}) at the
optimal scale $\mu_x^* = \mu_x^{\rm *,opt}$ truncated at $a_s^0$ (fine-dotted),
$a_s$ (dotted), $a_s^2$ (dashed double-dotted), $a_s^3$ (dashed dotted),
and $a_s^4$ (solid) are plotted as functions of $|x|$.
}
\label{fig:vv_lbest}
\end{center}
\end{figure}

The expansion at the scale $\mu_x^{\rm *,opt}$ reads
\begin{align}
\Pi_V^{\rm\overline{MS}}(x)\bigg|_{n_f=3}
= {6\over\pi^4x^6}\Big(
1&+a_s(\mu_x^{\rm *,opt}) +3.1431a_s(\mu_x^{\rm *,opt})^2
\notag\\
&+4.8432a_s(\mu_x^{\rm *,opt})^3
-33.819a_s(\mu_x^{\rm *,opt})^4
+O(a_s^5)
\Big).
\label{eq:vv_best_l_nf3}
\end{align}
As shown in Fig.~\ref{fig:vv_lbest}, the choice of $\mu_x^{\rm*,opt}$ shows
better convergence.
The gray region in the figure represents the higher order uncertainty,
which is estimated
by the maximum difference between
the correlator at $\mu_x^* = \mu_x^{*,\rm opt}$ and those at $\mu_x^*$
in $[{1\over2}\mu_x^{*,\rm opt},2\mu_x^{\rm*,opt}]$.

Next, we consider the scalar correlator.
The scalar channel is more complicated due to the scale dependence
of the scalar operator $S(x)$.
Using the beta function \cite{vanRitbergen:1997va} and the anomalous
dimension \cite{Chetyrkin:1997dh,Vermaseren:1997fq},
we can treat a general expression of the perturbative expansion of the scalar
correlator, which is written as
\begin{equation}
\Pi_S^{\rm\overline{MS}}(\mu_x';x)
= {3\over\pi^4x^6}\left(1+\sum_{i=1}^\infty
C_i^S(\mu_x^*,\mu_x')a_s(\mu^*_x)^i\right),
\label{eq:ss_corr_pt}
\end{equation}
where the first argument $\mu_x^*$ of the perturbative coefficients is the
renormalization scale of the strong coupling constant in the perturbative
series and the second $\mu_x'$ is the renormalization scale of the scalar
operator.
Chetyrkin and Maier \cite{Chetyrkin:2010dx} gave the perturbative
coefficients of correlators at the renormalization scale
$\mu_x^* = \mu_x' = \tilde\mu_x$, which
is $C_i^S(\tilde\mu_x,\tilde\mu_x)$ in our notation.
The coefficients $C_i^S(\mu_x,\mu_x)$ and $C_i^S(\tilde\mu_x,\tilde\mu_x)$
are summarized in Appendix~\ref{sec:app_pert}.

\begin{figure}[t]
\begin{center}
\includegraphics[width=120mm]{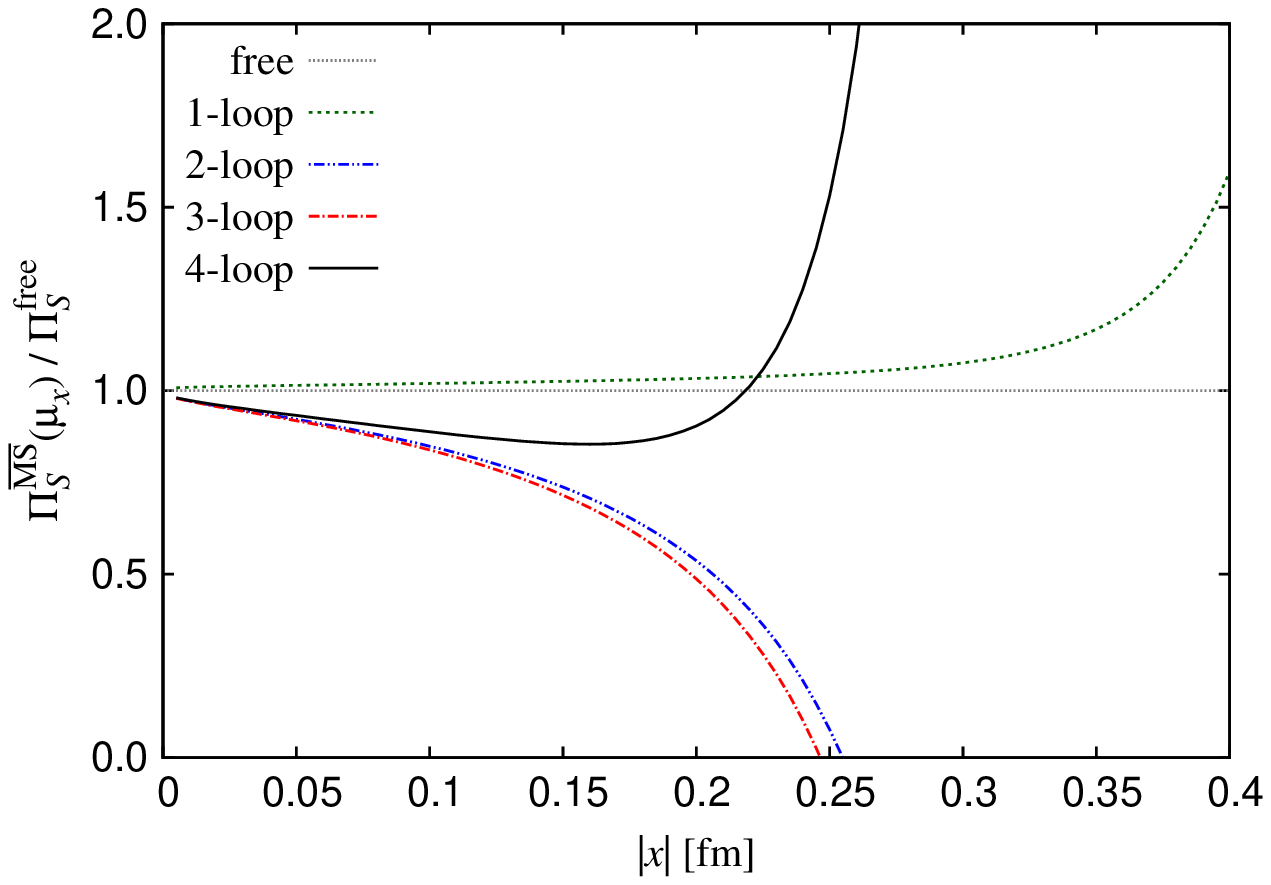}
\caption{
Scalar correlator renormalized at $\mu_x$ in $\rm\overline{MS}$ scheme
for $n_f = 3$.
The results at $\mu_x^* = \mu_x' = \mu_x$
truncated at $a_s^0$ (fine-dotted),
$a_s$ (dotted), $a_s^2$ (dashed double-dotted), $a_s^3$ (dashed dotted),
and $a_s^4$ (solid) are plotted.
}
\label{fig:ss_l0.0_t0.0_non_evlv}
\vspace{4.2mm}
\includegraphics[width=120mm]{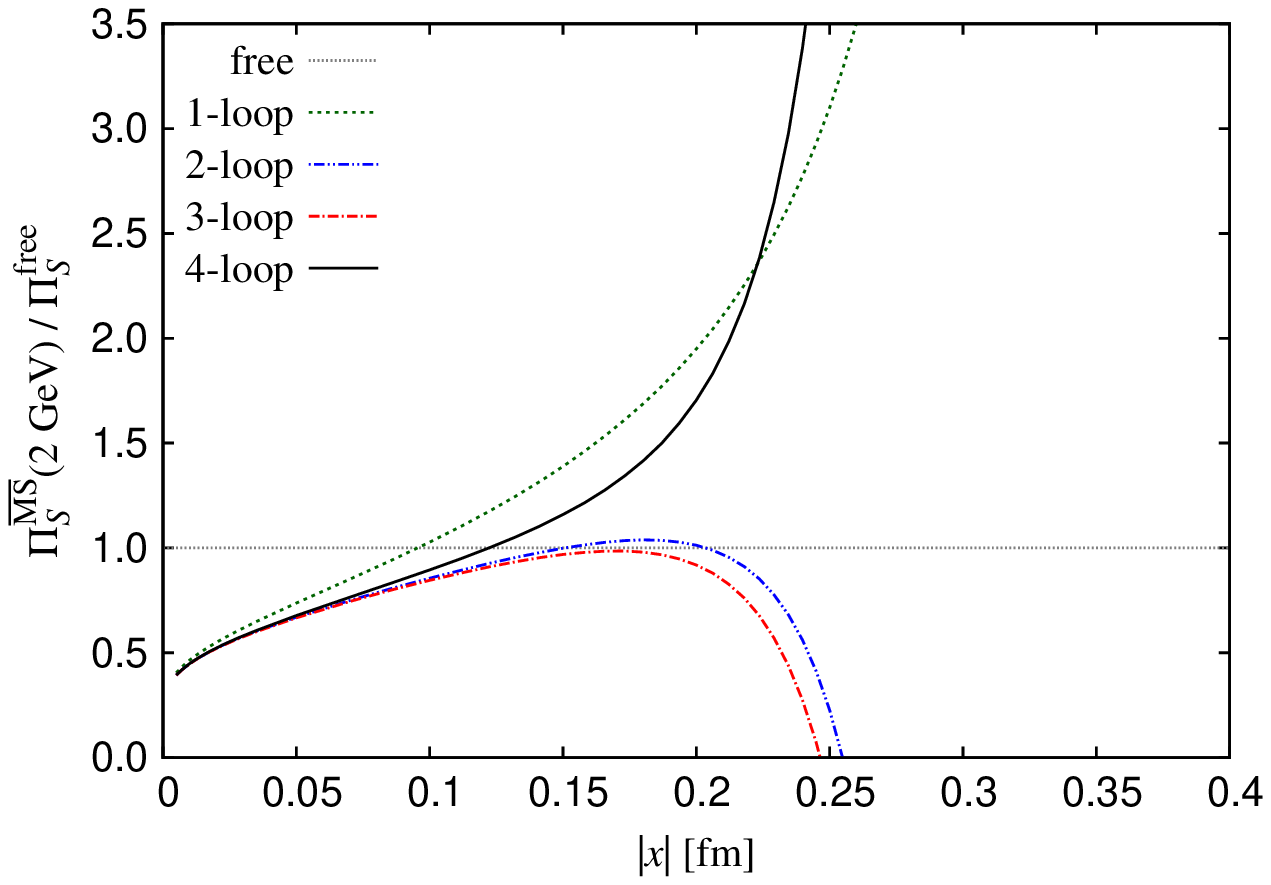}
\caption{
Scalar correlator renormalized at 2~GeV in $\rm\overline{MS}$ scheme
for $n_f = 3$, which is calculated from the perturbative
series at $\mu_x^* = \mu_x' = \mu_x$
and the scale evolution.
The results truncated at $a_s^0$ (fine-dotted),
$a_s$ (dotted), $a_s^2$ (dashed double-dotted), $a_s^3$ (dashed dotted),
and $a_s^4$ (solid) are plotted.
}
\label{fig:ss_l0.0_t0.0}
\end{center}
\end{figure}

Figure~\ref{fig:ss_l0.0_t0.0_non_evlv} shows the convergence of
the perturbative expansion at $\mu_x^* = \mu_x' = \mu_x$ with
$n_f=3$.
The correlator in the figure is normalized by the tree-level one,
$\Pi_S^{\rm free}(x) = 3/\pi^4x^6$.
The scalar channel also shows poor convergence.
The BLM scale is not directly applicable for scale dependent quantities.
In fact, the BLM scale of the scalar correlator renormalized at
$\mu_x$ is unstable, {\it i.e.}
$\mu_x^{\rm BLM} \simeq 8.8\mu_x$ for
$\Pi_S^{\rm\overline{MS}}(\tilde\mu_x;x)$
while
$\mu_x^{\rm BLM} \simeq 4\times10^3\mu_x$ for
$\Pi_S^{\rm\overline{MS}}(\mu_x;x)$.

Since the purpose of this work is to determine the renormalization constant
at 2~GeV in the $\rm\overline{MS}$ scheme, we 
perform the scale evolution of the scalar correlator (\ref{eq:ss_corr_pt})
from $\mu_x'$ to 2~GeV by a numerical integral of the mass anomalous
dimension \cite{Chetyrkin:1997dh,Vermaseren:1997fq}.
Figure~\ref{fig:ss_l0.0_t0.0} shows the scalar correlator
calculated from the perturbative series
at $\mu_x^* = \mu_x' = \mu_x$ and the scale evolution with $n_f=3$.
This calculation is convergent only below $|x| \sim 0.06$~fm.

\begin{figure}[t]
\begin{center}
\subfigure{\mbox{\raisebox{1mm}{\includegraphics[width=81mm]{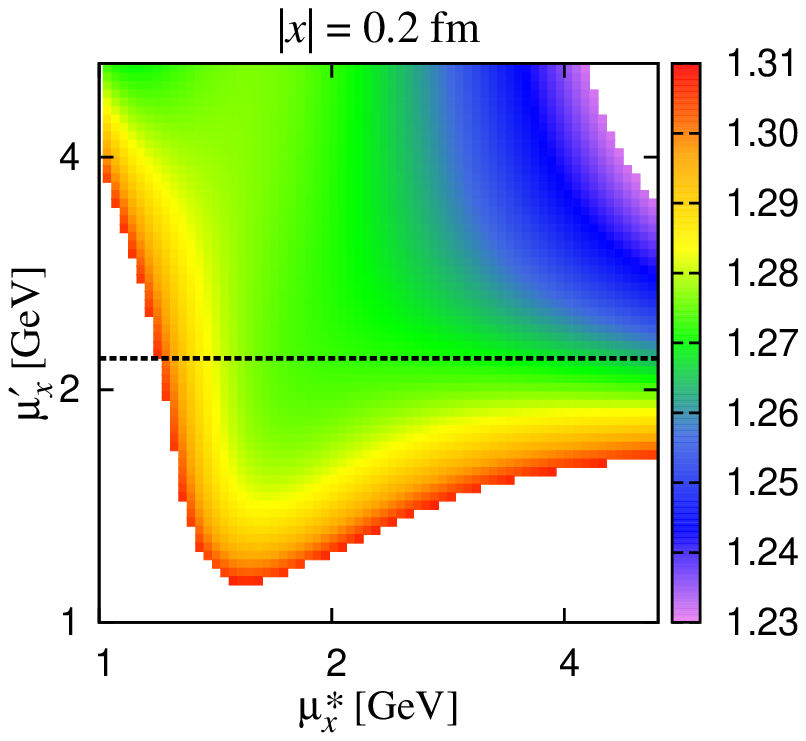}}}}
\subfigure{\mbox{\raisebox{1mm}{\includegraphics[width=81mm]{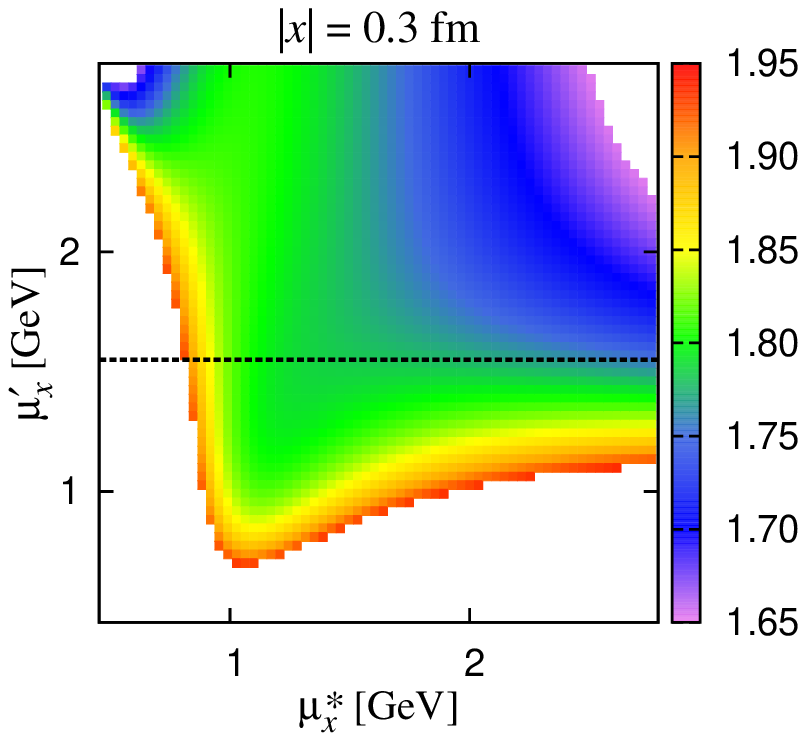}}}}
\subfigure{\mbox{\raisebox{1mm}{\includegraphics[width=81mm]{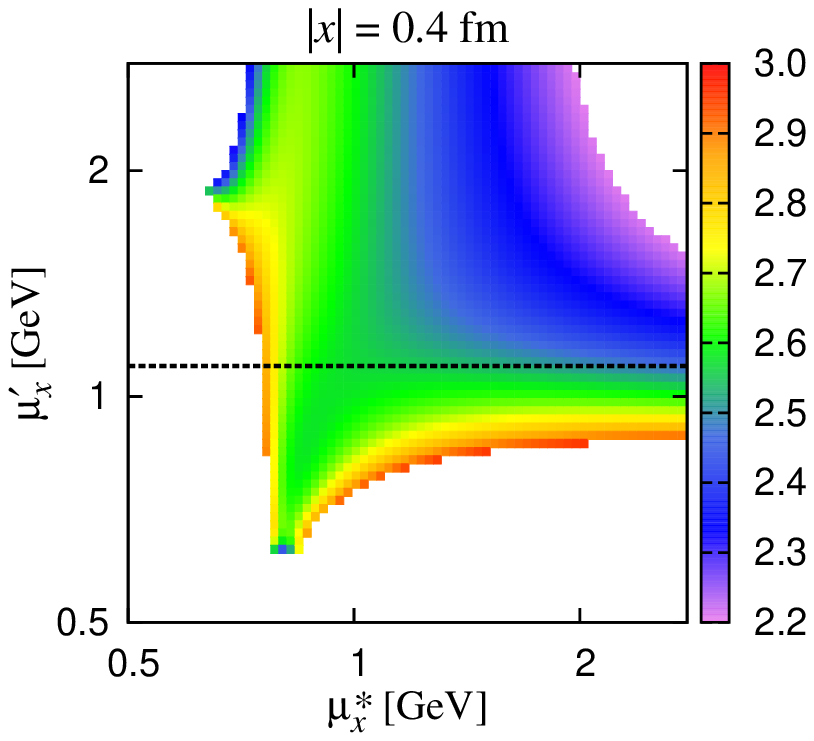}}}}
\subfigure{\mbox{\raisebox{1mm}{\includegraphics[width=81mm]{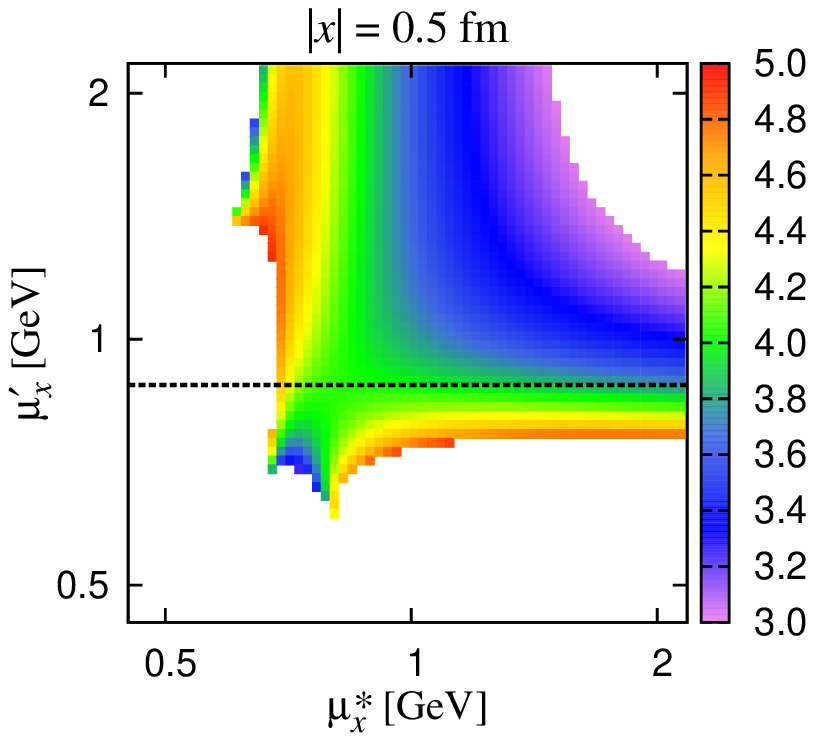}}}}
\caption{
$\Pi_S^{\rm\overline{MS}}(2{\rm\ GeV};x)/\Pi_S^{\rm free}(x)$ at
the four-loop level at $n_f = 3$.
The results at the specific distances 0.2~fm (top/left), 0.3~fm (top/right),
0.4~fm (bottom/left), and 0.5~fm (bottom/right) are shown as functions
of $\mu^*_x$ and $\mu'_x$.
The dashed lines stand for our choice of $\mu_x'$, where the correlator
shows small sensitivity to $\mu'_x$ and $\mu_x^*$.
}
\label{fig:ltdps_ss_0.20fm}
\end{center}
\end{figure}

Although the scalar correlator
$\Pi_S^{\rm\overline{MS}}({2\rm\ GeV};x)$ after the scale
evolution has to be independent of both the scales $\mu^*_x$
and $\mu'_x$, finite order calculations may depend on them.
Figure~\ref{fig:ltdps_ss_0.20fm} shows the dependences on $\mu_x^*$ and $\mu_x'$
of the four-loop results with $n_f=3$ at four representative distances renormalized at 2~GeV in
the $\rm\overline{MS}$ scheme.
To choose optimal values of $\mu_x^*$ and $\mu_x'$, we focus on the region
with mild dependence of the correlator.
We choose the optimal values of $\mu'_x (= \mu_x'^{\rm\ opt})$ as indicated
by the dashed lines in Fig.~\ref{fig:ltdps_ss_0.20fm}.
On these lines, the correlator depends on $\mu_x^*$ mildly
and the dependence on $\mu_x'$ is also relatively small.
Numerically, the choice is
\begin{equation}
\mu_x'^{\rm\ opt} = \e^{0.8} \mu_x
\simeq {2.2\over x}.
\label{eq:mup_cond}
\end{equation}

\begin{figure}[t]
\begin{center}
\subfigure{\mbox{\raisebox{1mm}{\includegraphics[width=81mm]{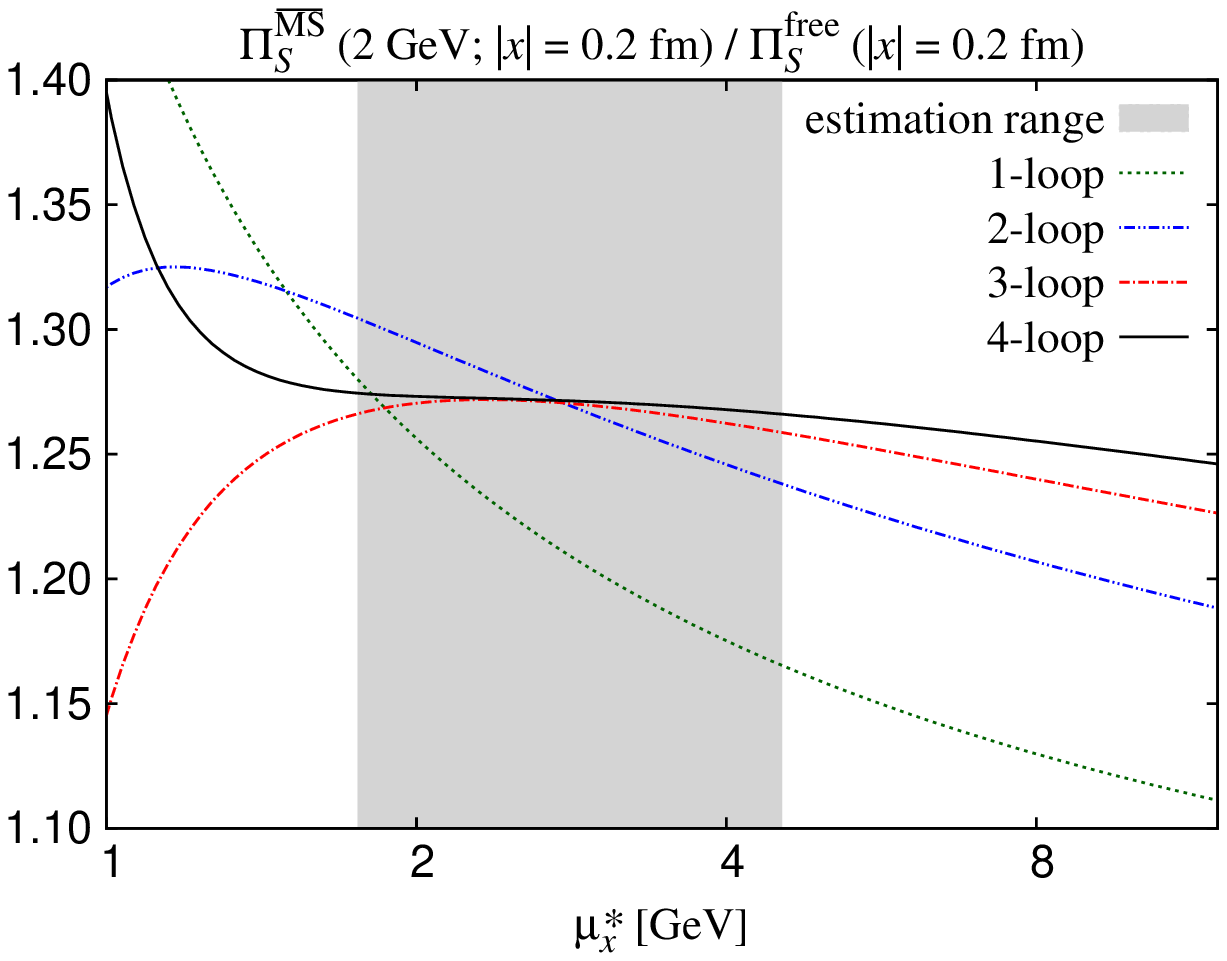}}}}
\subfigure{\mbox{\raisebox{1mm}{\includegraphics[width=81mm]{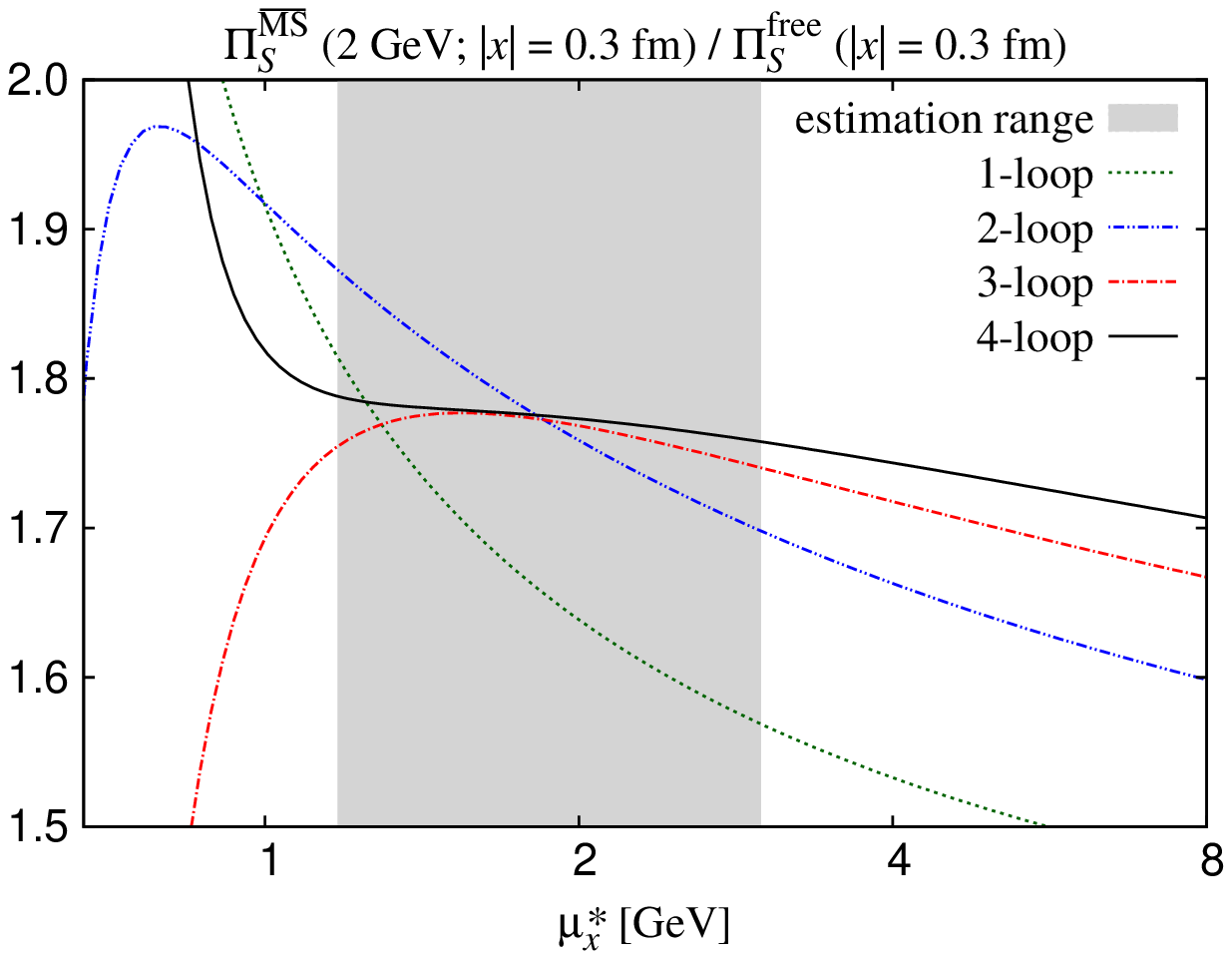}}}}
\subfigure{\mbox{\raisebox{1mm}{\includegraphics[width=81mm]{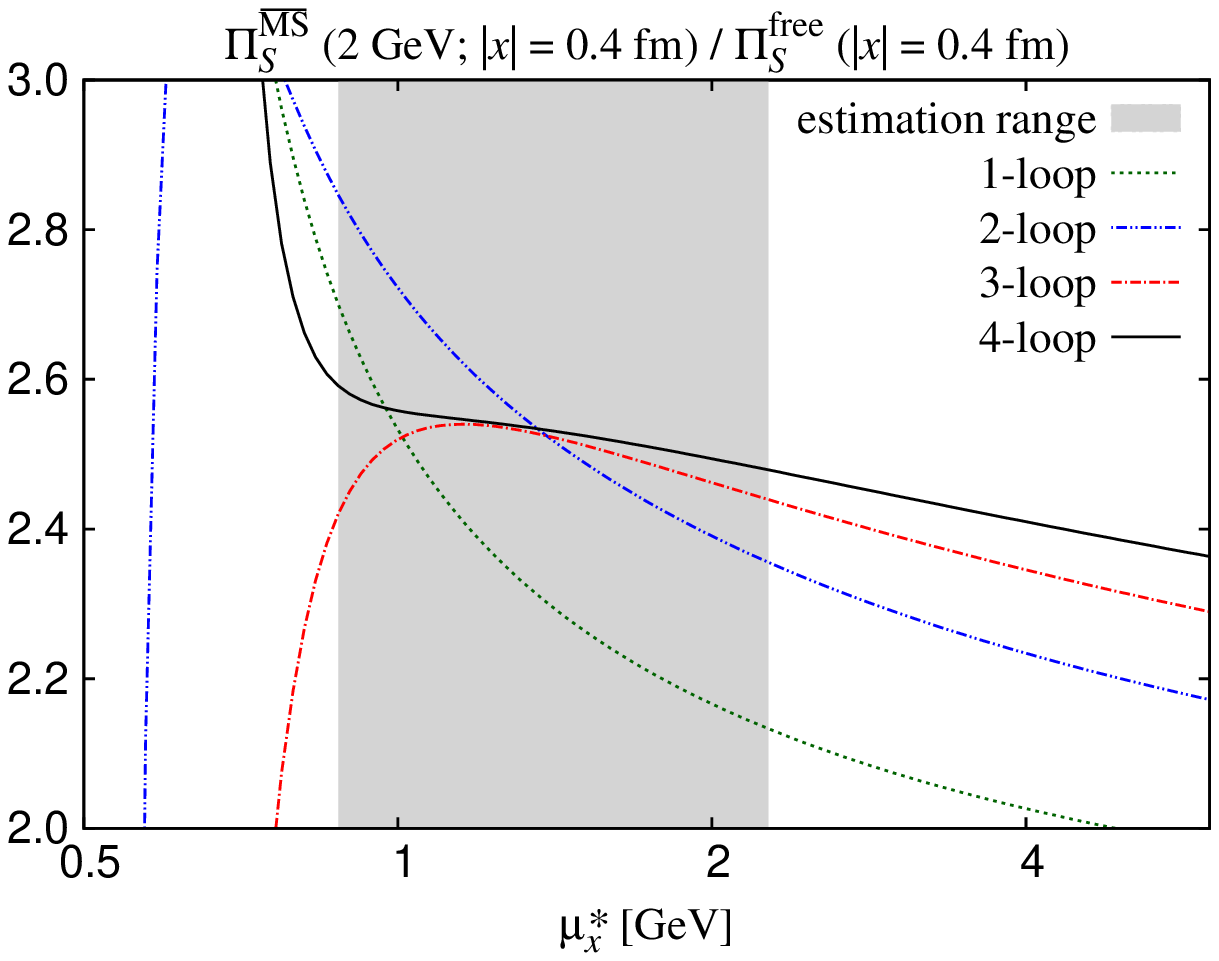}}}}
\subfigure{\mbox{\raisebox{1mm}{\includegraphics[width=81mm]{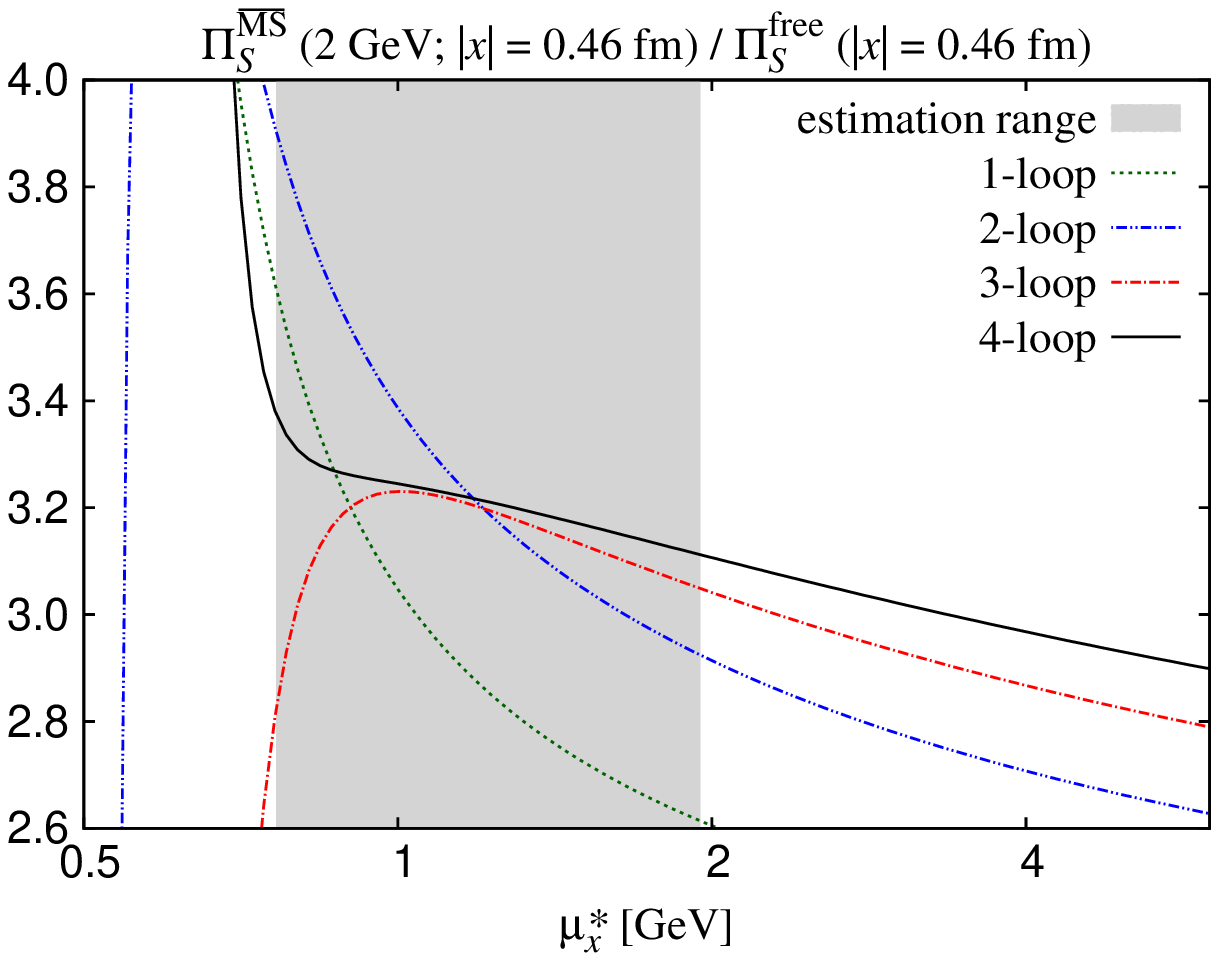}}}}
\caption{
Scalar correlator renormalized at 2~GeV in the $\rm\overline{MS}$ scheme
at the specific distances 0.2~fm (top/left), 0.3~fm (top/right),
0.4~fm (bottom/left), and 0.46~fm (bottom/right) as functions
of $\mu^*_x$.
They are calculated at $n_f = 3$ and $\mu_x'$ is set by (\ref{eq:mup_cond}).
The results truncated at
$a_s$ (dotted), $a_s^2$ (dashed double-dotted), $a_s^3$ (dashed dotted),
and $a_s^4$ (solid) are plotted.
The gray band represents the region in which we estimate the uncertainty
of the perturbative calculation.
}
\label{fig:ldps_ss_tbest}
\end{center}
\end{figure}

The detailed dependence of the scalar correlator on $\mu_x^*$
at $\mu_x' = \mu_x'^{\rm\ opt}$ is shown in
Fig.~\ref{fig:ldps_ss_tbest} for several distances in 0.2--0.46~fm,
in which we determine the renormalization factor of the scalar operator.
These figures show results at each loop order up to the four-loop level.
We then choose the optimal value $\mu_x^{*,\rm opt}$ of
the scale $\mu_x^*$ as the value which minimizes the
dependence on $\mu_x^*$,
\begin{equation}
\mu_x^{*,\rm opt} = \e^{1.05}\mu_x
\simeq {2.9\over x}.
\label{eq:ss_best_l}
\end{equation}

We estimate the uncertainty by varying $\mu^*_x$ in a region including
$\mu_x^{*,\rm opt}$ in the middle as we did for the vector channel.
Since the coupling constant blows up as $\mu_x^*$ approaches $\Lambda_{\rm QCD}$,
we need to avoid too small $\mu_x^*$.
Therefore, our choice is $[{1\over1.6}\mu_x^{*,\rm opt},1.6\mu_x^{*,\rm opt}]$
in order not to use the coupling constant at the scale smaller than $0.75$~GeV.
The region is shown by the gray band in
Fig.~\ref{fig:ldps_ss_tbest}.

\begin{figure}[t]
\begin{center}
\includegraphics[width=120mm]{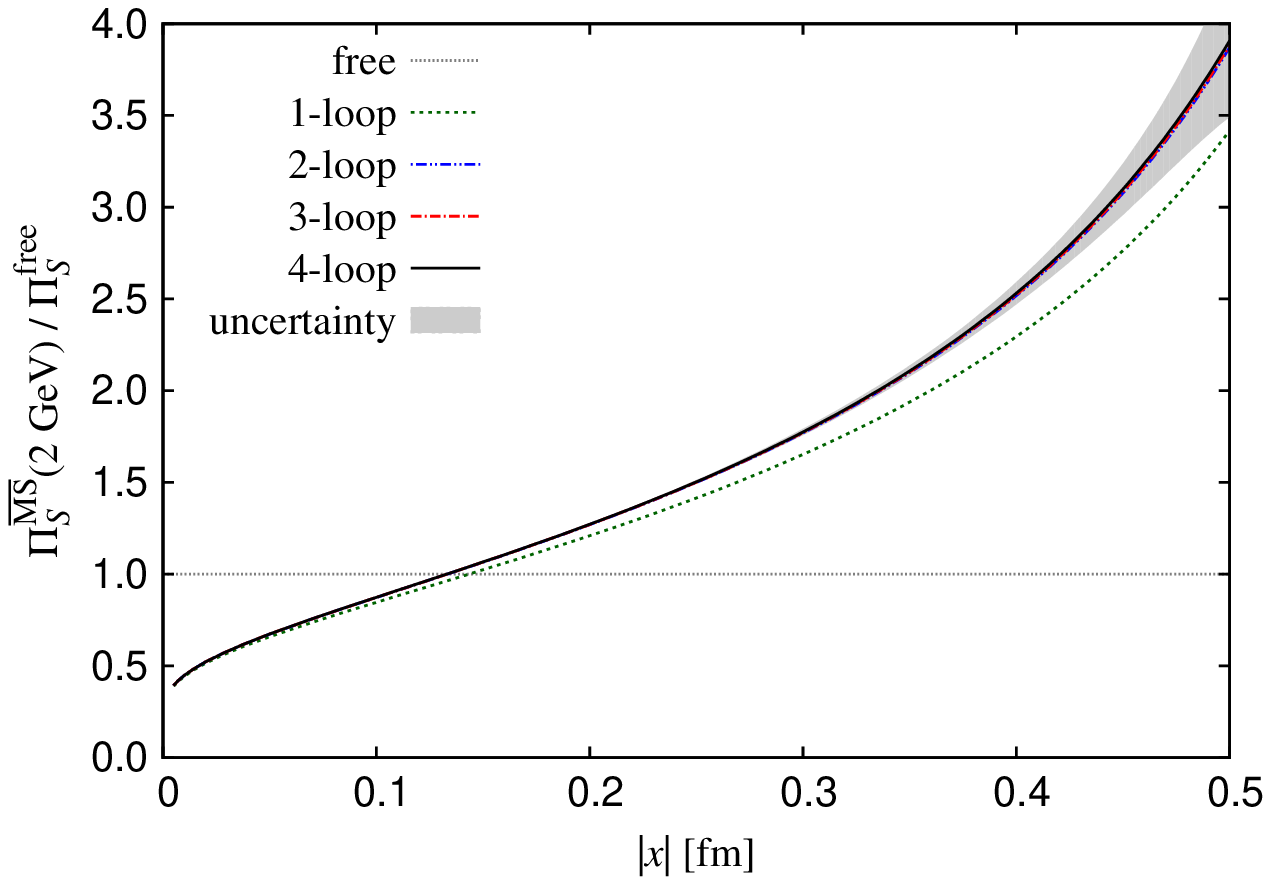}
\caption{
Scalar correlator renormalized at 2~GeV in the $\rm\overline{MS}$ scheme
at $n_f = 3$.
The scale parameters $\mu_x'$ and $\mu_x^*$ are set by (\ref{eq:mup_cond})
and (\ref{eq:ss_best_l}).
The results truncated at $a_s^0$ (fine-dotted),
$a_s$ (dotted), $a_s^2$ (dashed double-dotted), $a_s^3$ (dashed dotted),
and $a_s^4$ (solid) are plotted.
The gray region stands for the uncertainty of higher order corrections,
which is estimated by the sensitivity to $\mu_x^*$.
}
\label{fig:ss_ltbest}
\end{center}
\end{figure}

Setting $\mu_x'$ and $\mu_x^*$ by (\ref{eq:mup_cond}) and (\ref{eq:ss_best_l}),
we obtain the following numerical expansion at $n_f = 3$,
\begin{align}
\Pi_S^{\rm\overline{MS}}(\mu_x'^{\rm\ opt};x)\Big|_{n_f=3}
= {3\over\pi^4x^6}\Big(
1 &+ 3.4029a_s(\mu_x^{*,\rm opt}) +9.7142a_s(\mu_x^{*,\rm opt})^2
\notag\\*
&+1.7011a_s(\mu_x^{*,\rm opt})^3 +26.366a_s(\mu_x^{*,\rm opt})^4
+ O(a_s^5)
\Big).
\label{eq:ss_ltbest}
\end{align}
The four-loop correction in this series is much smaller than that
$\sim579 a_s(\mu_x)^4$ in the perturbative series at $\mu_x^* = \mu_x$
(see Appendix~\ref{sec:app_pert}),
implying that the convergence of the perturbative series (\ref{eq:ss_ltbest})
is better than that at $\mu_x^* = \mu_x$.
In fact, evolving the renormalization scale of the correlator to 2~GeV, we
obtain a well convergent correlator as shown in Fig.~\ref{fig:ss_ltbest}.
The figure also shows the uncertainty of the perturbative calculation
by the gray band, which is estimated by the maximum difference between
the correlator at $\mu_x^* = \mu_x^{*,\rm opt}$ and those at $\mu_x^*$
in $[{1\over1.6}\mu_x^{*,\rm opt},1.6\mu_x^{\rm*,opt}]$.

\subsection{Operator Product Expansion}
\label{subsec:OPEforms}

In the previous subsection, we discuss the correlators in perturbation theory.
In the real QCD vacuum with the spontaneous breaking of chiral symmetry,
however, additional terms with vacuum expectation values of operators may appear,
inducing the violation of the degeneracy between the scalar and pseudoscalar,
and between the vector and axial-vector channels.

In the momentum space, the Operator Product Expansion (OPE)
of correlators is sketched as
\begin{equation}
\widetilde\Pi_\Gamma^{\rm OPE}(q^2)
= \widetilde\Pi_\Gamma^{\rm pert}(q^2)
+ \sum_{\cal O}\tilde c_{\cal O}^\Gamma(q^2)q^{2-\dim\cal O}\langle{\cal O}\rangle,
\end{equation}
where the first term on the RHS corresponds to the perturbative contribution
including the quark mass corrections.
Here $\cal O$ denotes operators with the mass dimension four and higher,
\begin{equation}
{\cal O} \in \{m_q\bar qq, a_sG^2, m_q g_s\bar qGq, a_s(\bar qq)^2, \cdots\},
\end{equation}
and the Wilson coefficients $\tilde c_{\cal O}^\Gamma(q^2)$ do not have
mass dimensions.
In general, $\widetilde c_{\cal O}^\Gamma$'s depend on
$q^2$ logarithmically as $\ln(q^2/\mu^2)$ and on the quark mass as $m_q^2/q^2$.
These coefficients are known up to
$\dim{\cal O} = 8$ \cite{Shifman:1978bx,Broadhurst:1985js,Jamin:1992se}.

The short-distance behavior of the correlators in the coordinate space is obtained
by Fourier transform
\begin{equation}
\Pi_\Gamma^{\rm OPE}(x^2)
= \int{\td^4q\over(2\pi)^4}\e^{\img qx}\widetilde \Pi_\Gamma^{\rm OPE}(q^2)
= \Pi_\Gamma^{\rm pert}(x^2)
+ \sum_{\cal O}c_{\cal O}^\Gamma(x^2)x^{\dim{\cal O} - 6}\langle{\cal O}\rangle.
\label{eq:FT_OPE}
\end{equation}
A convenient set of relations between the momentum space and
the coordinate space
is summarized in \cite{Narison:2001ix}.
Using these relations, we obtain the Wilson coefficients in the coordinate space:
\begin{align}
&c_{m\bar qq}^{S/P}
= {1\pm2\over4\pi^2}
+ O(a_s,m_q^2x^2),
\ \ 
c_{m\bar qq}^{V/A}
= {1\mp4\over2\pi^2}
+ O(a_s,m_q^2x^2),
\\
&c_{a_sG^2}^{S/P}
= {1\over32\pi^2}
+ O(a_s,m_q^2x^2),
\ \ 
c_{a_sG^2}^{V/A}
= - {1\over16\pi^2}
+ O(a_s,m_q^2x^2),
\label{eq:c_G2}
\\
&c_{mg_s\bar qGq}^{S/P}
= \mp {1\over16\pi^2}\ln (x/x_0)^2 + O(a_s,m_q^2x^2),
\ \ 
c_{mg_s\bar qGq}^{V/A} = 0,
\label{eq:c_mqGq}
\\
&c_{a_s(\bar qq)^2}^{S/P}
= {4\pm11\over27}\ln (x/x_0)^2 + O(a_s,m_q^2x^2),
\ \ 
c_{a_s(\bar qq)^2}^{V/A}
= {2(2\mp9)\over27}\ln (x/x_0)^2 + O(a_s,m_q^2x^2).
\label{eq:c_qq2}
\end{align}
Here, the coefficients of the four-quark condensate $c_{a_s(\bar qq)^2}^\Gamma$
are calculated using the vacuum saturation approximation \cite{Narison:2001ix}.
The logarithmic contributions in (\ref{eq:c_mqGq}) and (\ref{eq:c_qq2})
involve dimensionful variable $x_0$,
which appears from divergences of the integral defining the Fourier transform and
are regularization and scheme dependent.

The perturbative contribution on the RHS of (\ref{eq:FT_OPE})
also includes the mass term which also violates the degeneracy of
the scalar and pseudoscalar, or the vector and axial-vector channels.
The massless part is already discussed in the previous subsection.
The mass term contributes as
\begin{align}
\Pi_{P/S}^{\rm pert}(x^2)
- \Pi_{P/S}^{\rm pert, massless}(x^2) &=
- {3(2\pm1)m_q^2\over4\pi^4x^4}
+ O(\alpha_s/\pi,m_q^4/x^2),
\\
\Pi_{V/A}^{\rm pert}(x^2)
- \Pi_{V/A}^{\rm pert, massless}(x^2) &=
- {6(1\mp1)m_q^2\over\pi^4x^4}
+ O(\alpha_s/\pi,m_q^4/x^2).
\end{align}
In the renormalization condition (\ref{eq:zfactor_cond}), we add
these mass correction terms to $\Pi_\Gamma^{\rm\overline{MS}}({2\rm\ GeV};x)$.

\section{Lattice calculation}
\label{sec:lat_calc}

\subsection{Lattice Setup}
\label{subsec:ensembles}

\tabcolsep = 6pt
\begin{table}[htb]
\caption{
Lattice ensembles used in this work.
}
\label{tab:ensembles}
\begin{center}
\begin{tabular}{cccccccc}
\hline
\hline
$\beta$ & $a$ [fm] & $N_s^3\times N_t\times L_s$
& $am_s$ & $am_q$ & $aM_\pi$ & $N_{\rm conf}$ & $N_{\rm src}$\\
\hline
4.17 & 0.0804 & $32^3\times64\times12$ & 0.0300 & 0.0070 & 0.1263(4) & 200 & 4\\
& & & & 0.0120 & 0.1618(3) & 200 & 2 \\
& & & & 0.0190 & 0.2030(3) & 200 & 2 \\
& & $48^3\times96\times12$ & 0.0400 & 0.0035 & 0.0921(1) & 200 & 2 \\
& & $32^3\times64\times12$ & & 0.0070 & 0.1260(4) & 200 & 4 \\
& & & & 0.0120 & 0.1627(3) & 200 & 2 \\
& & & & 0.0190 & 0.2033(3) & 200 & 2 \\
\hline
4.35 & 0.0547 & $48^3\times96\times8$ & 0.0180 & 0.0042 & 0.0820(3) & 200 & 2 \\
& & & & 0.0080 & 0.1127(3) & 200 & 1 \\
& & & & 0.0120 & 0.1381(3) & 200 & 1 \\
& & & 0.0250 & 0.0042 & 0.0831(4) & 200 & 2 \\
& & & & 0.0080 & 0.1130(3) & 200 & 1 \\
& & & & 0.0120 & 0.1387(3) & 200 & 1 \\
\hline
4.47 & 0.0439 & $64^3\times128\times8$ & 0.0150 & 0.0030 & 0.0632(2) & 200 & 1 \\
\hline
\hline
\end{tabular}
\end{center}
\end{table}

In this work, we perform lattice simulations with $2+1$-flavor dynamical M\"obius
domain-wall fermions \cite{Brower:2004xi,Brower:2012vk} with three-step stout
link smearing \cite{Morningstar:2003gk} and the tree-level Symanzik improved
gauge action \cite{Luscher:1984xn}.
The properties of the gauge ensembles used in this analysis are summarized in
Table~\ref{tab:ensembles}.
The input strange quark mass $m_s$ is only for the sea quark,
while the mass $m_q$ of two degenerate quarks, up and down, is used
for both the valence and sea quarks.
The computed pion masses $M_\pi$ are in the region 230--500~MeV.

For each ensemble, $N_{\rm conf}=200$ configurations are sampled from 10,000
molecular dynamics time.
For each configuration, we calculate correlators from one or more ($N_{\rm src}$)
source points.
We use the IroIro$++$ simulation code \cite{Cossu:2013ola} for these calculations.

Considering the violation of rotational symmetry,
we distinguish different lattice points that are not related by
$90^\circ$ rotations in the four-dimensional cubic group
and average the correlators on the lattice over the lattice points
that are related by $90^\circ$ rotations.
We then have 322 sets of different separations in the region
$1 \le (x/a)^2 \le 100$.

\subsection{Reduction of discretization effect}
\label{subsec:free_correc}

\begin{figure}[t]
\vspace{-2.2mm}
\begin{center}
\includegraphics[width=120mm]{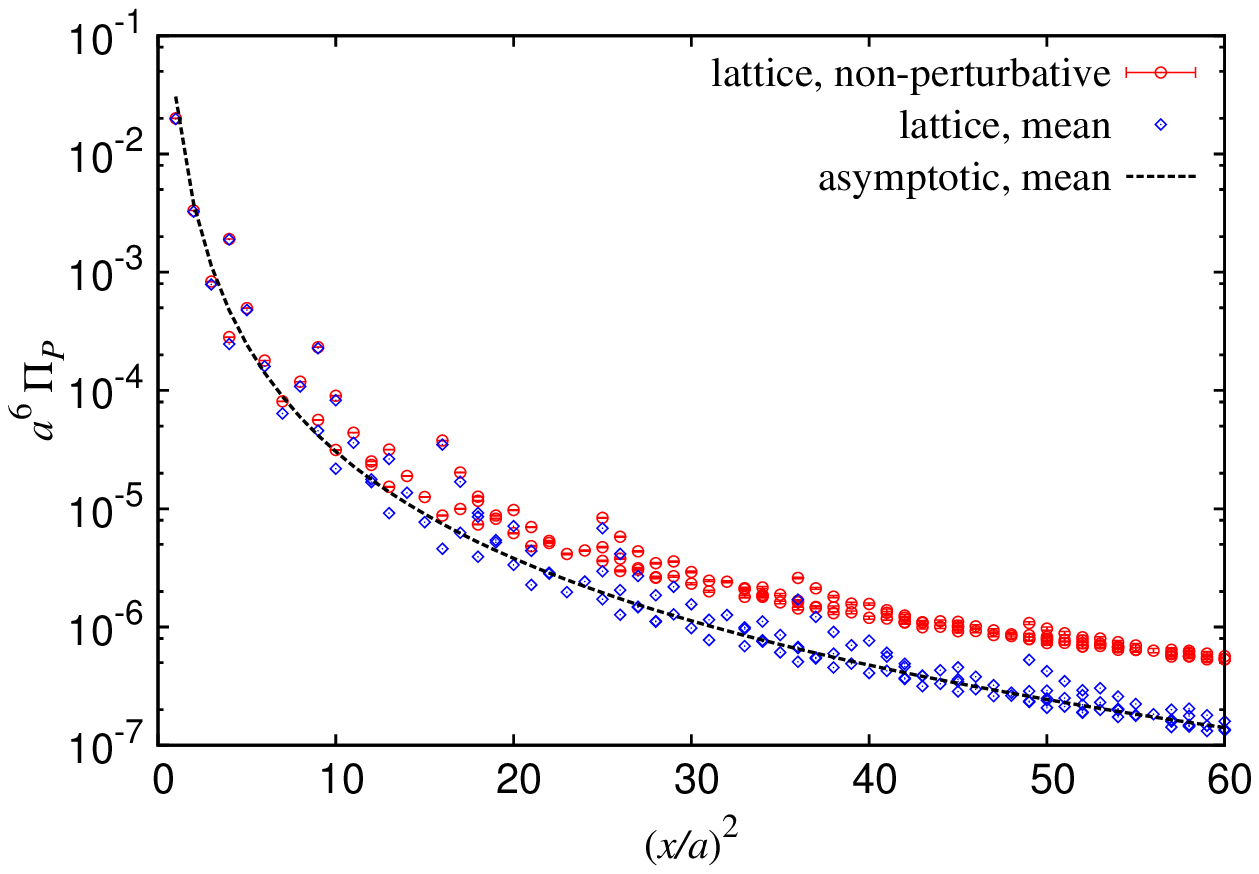}
\caption{
Pseudoscalar correlator obtained by the lattice calculation (circles)
and its mean field approximation (diamonds) at the same valence mass.
The dashed curve represents the asymptotic behavior of the mean field
approximation.
The lattice data on the $48^3\times96$ at $\beta = 4.35$ and
$(am_q,am_s) = (0.0042,0.0180)$ are plotted as a representative.
}
\label{fig:x2vspp_raw}
\end{center}
\begin{center}
\includegraphics[width=120mm]{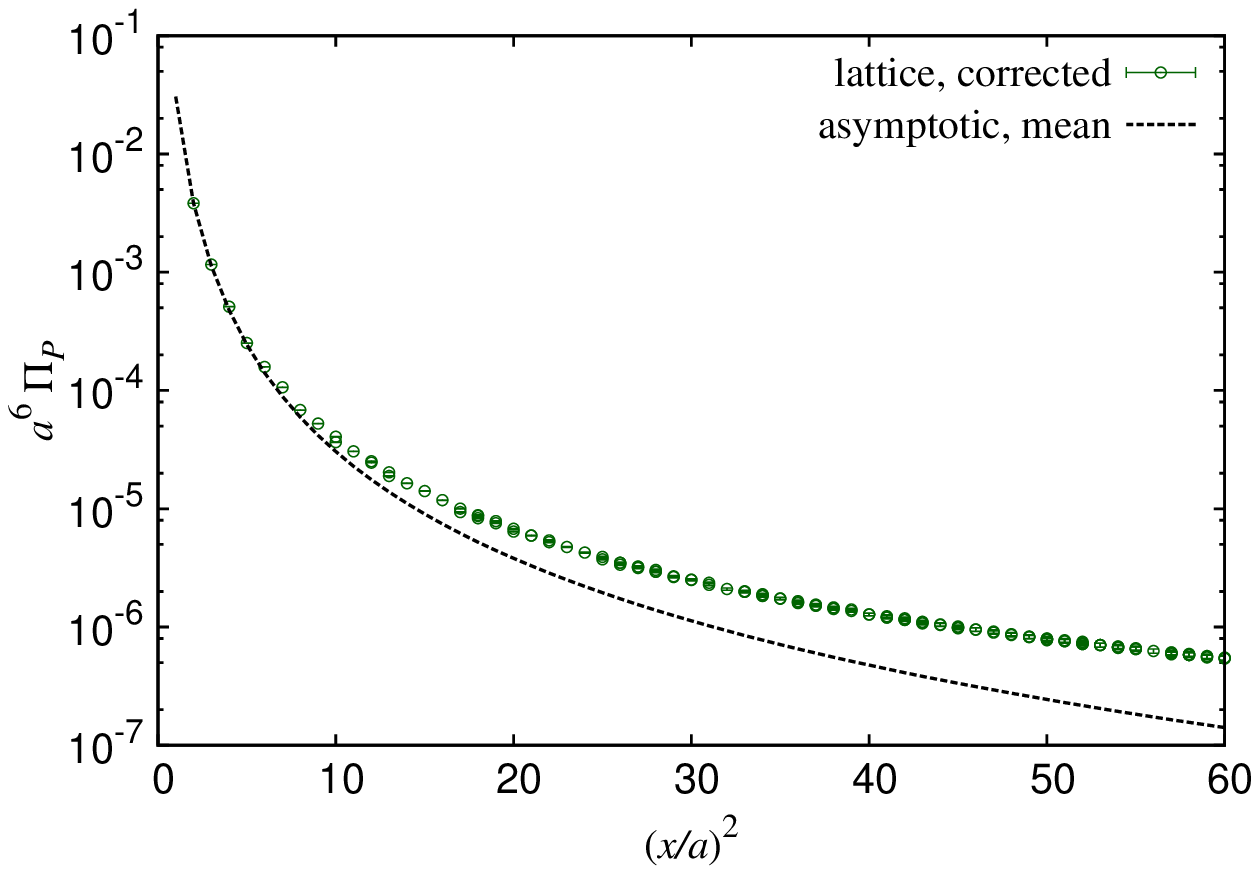}
\caption{
Pseudoscalar correlator after applying the subtraction
(\ref{eq:subt_disc_eff}).
The result on the same ensemble as in Fig.~\ref{fig:x2vspp_raw} is shown.
}
\label{fig:x2vspp_fcorrec}
\end{center}
\end{figure}

The pseudoscalar correlator $\Pi_P^{\rm lat}(x)$ calculated non-perturbatively
on the ensemble at $\beta = 4.35, (am_q, am_s) = (0.0042, 0.0180)$ is plotted
in Fig.~\ref{fig:x2vspp_raw}.
This figure also shows
the mean field approximation $\Pi_P^{\rm lat,mean}(x)$ of the
pseudoscalar correlator on the lattice
and its asymptotic form $\Pi_P^{\rm asym,mean}(x)$ in the long-distance limit.
The mean field approximation on the lattice is calculated by a contraction
of the propagators of the domain-wall fermions in the mean field theory
and its long-distance limit is calculated by applying Taylor expansion
(see Appendix~\ref{sec:free_corr} for more detail).

The non-perturbative lattice data are not on a smooth curve due to discretization
effects.
However, the similar discretization effect as seen in the
free theory as well as in the mean field approximation describes 
the bulk of the discretization effects.
We therefore improve the lattice data by applying a subtraction,
\begin{equation}
\Pi_\Gamma^{\rm lat}(x) \rightarrow
\Pi_\Gamma^{\rm lat}(x)
- \big(\Pi_\Gamma^{\rm lat,mean}(x) - \Pi_\Gamma^{\rm asym,mean}(x)\big).
\label{eq:subt_disc_eff}
\end{equation}
As shown in Fig.~\ref{fig:x2vspp_fcorrec}, we obtain much smoother correlators.

\begin{figure}[t]
\begin{center}
\includegraphics[width=120mm]{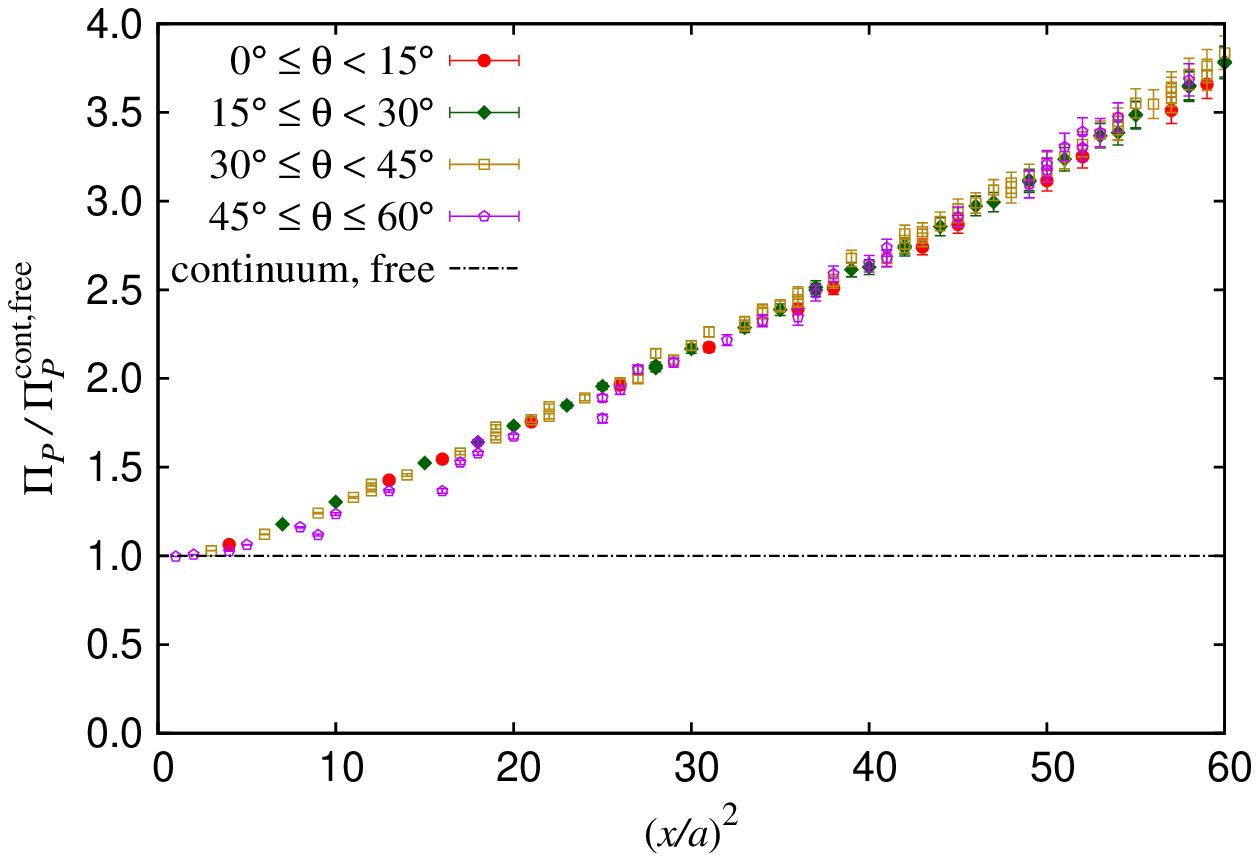}
\caption{
Pseudoscalar correlator divided by the tree-level continuum correlator
after applying the subtraction (\ref{eq:subt_disc_eff}).
The data in
$0^\circ \le \theta < 15^\circ$ (circles),
$15^\circ \le \theta < 30^\circ$ (diamonds),
$30^\circ \le \theta < 45^\circ$ (squares),
and $45^\circ \le \theta \le 60^\circ$ (pentagons) are separately plotted.
The result on the same ensemble as in Fig.~\ref{fig:x2vspp_raw} is shown.
}
\label{fig:x2vsppx6}
\end{center}
\vspace{3mm}
\begin{center}
\includegraphics[width=120mm]{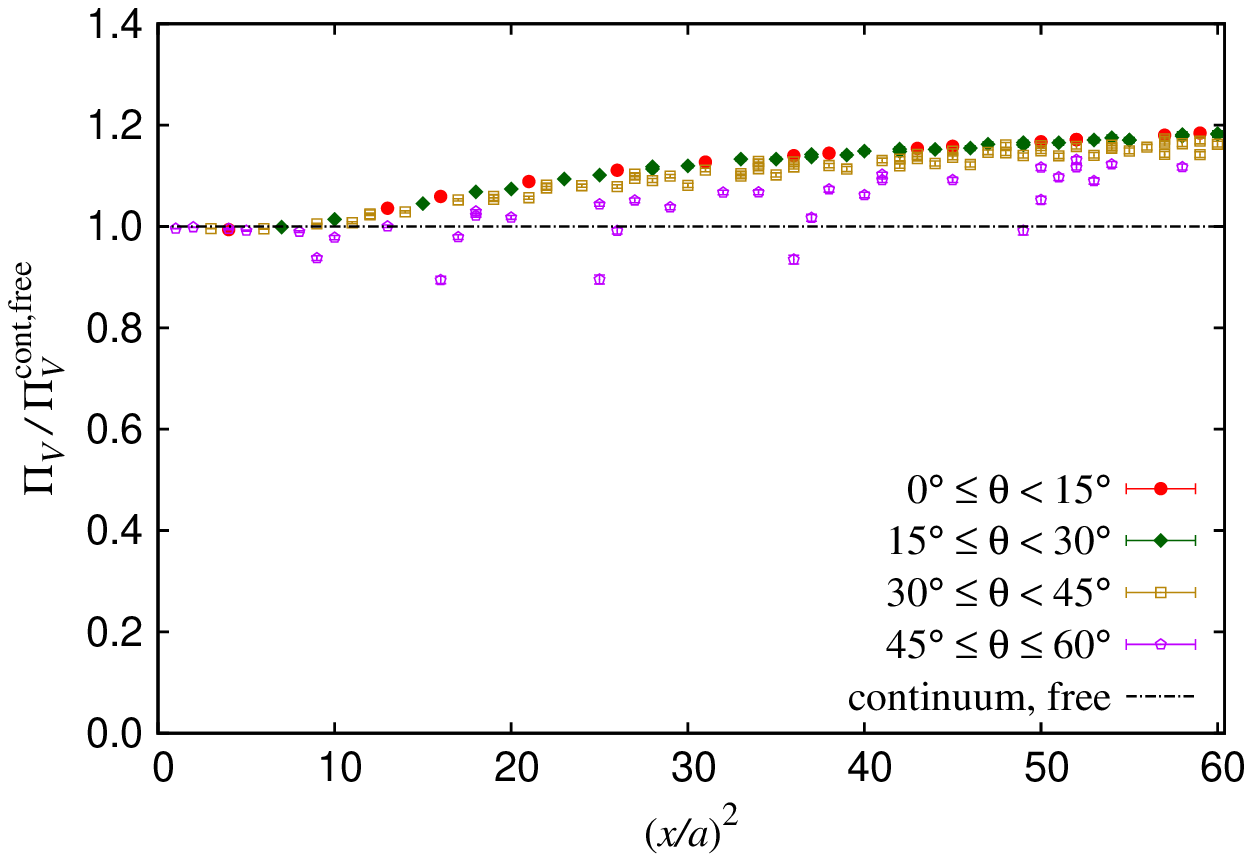}
\caption{
Same as Fig.~\ref{fig:x2vsppx6} but for the vector channel.
}
\label{fig:x2vsvvx6}
\end{center}
\end{figure}

\begin{figure}[t]
\begin{center}
\includegraphics[width=120mm]{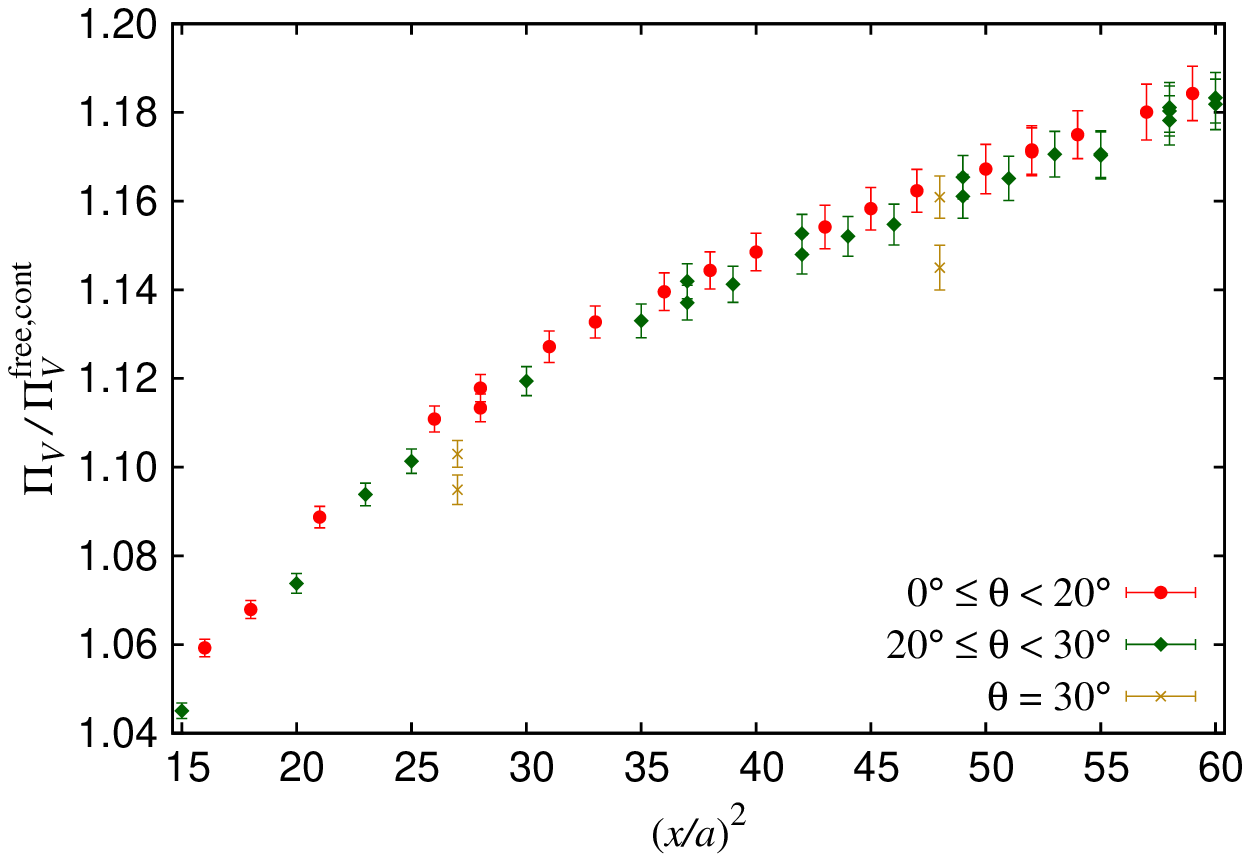}
\caption{
Detailed view of Fig.~\ref{fig:x2vsvvx6}.
The Data in
$0^\circ \le \theta < 20^\circ$ (circles),
$20^\circ \le \theta < 30^\circ$ (diamonds),
and at $\theta = 30^\circ$ (crosses) are separately plotted.
Data at $\theta > 30^\circ$ are omitted.
}
\label{fig:x2vsvvx6_zoomed}
\end{center}
\end{figure}

Figures~\ref{fig:x2vsppx6} and \ref{fig:x2vsvvx6} show the pseudoscalar and vector
correlators, respectively, after applying the subtraction (\ref{eq:subt_disc_eff}).
They are normalized by the continuum free correlator $\Pi_{P/V}^{\rm cont,free}(x)$
and plotted in a linear scale.
Here, we introduce a parameter $\theta$, which is defined as an angle
between the four-dimensional point $x$ and the direction $(1,1,1,1)$.
In four-dimension, $\theta$ is not larger than $60^\circ$.
This parameter is strongly correlated with the discretization effects
as discussed in \cite{Chu:1993cn,Cichy:2012is}, {\it i.e.} the discretization effects
increase as $\theta$ increases.
It is observed even after applying the subtraction (\ref{eq:subt_disc_eff}).

As seen in Fig.~\ref{fig:x2vsvvx6_zoomed}, which is a magnification of
Fig.~\ref{fig:x2vsvvx6} in the range $0^\circ \le \theta \le 30^\circ$,
there are discretization effects visible already at $\theta = 30^\circ$.
Although the points (0,3,3,3) and (1,1,3,4) are both at $(x/a)^2 = 27$ and
$\theta = 30^\circ$, the values disagree beyond the statistical error.
The same is observed for the data at (0,4,4,4) and (2,2,2,6) sharing
$(x/a)^2 = 48$ and $\theta = 30^\circ$.
The similar situation occurs more frequently for $\theta > 30^\circ$.
Namely, the data in the region $\theta \ge 30^\circ$ cannot
be simply parametrized by any functions of $\theta$, and we therefore omit them
in the analysis.
Figure~\ref{fig:x2vsvvx6_zoomed} also indicates that the
lattice data at $\theta < 30^\circ$ may slightly depend on $\theta$.
For the determination of the renormalization factor in the next section,
we separately treat the discretization effects at
$\theta < 20^\circ$ and at $\theta \ge 20^\circ$.

\subsection{Subtraction of finite volume effect}
\label{subsec:FVE}

Point-to-point correlators may contain finite volume effect,
which relates correlators $\Pi_\Gamma^{L^3\times T}(x)$ in a
finite volume with periodic boundaries
to those $\Pi_\Gamma^\infty(x)$ in the infinite volume as
\begin{equation}
\Pi_\Gamma^{L^3\times T} (x)
= \Pi_\Gamma^\infty(x) + \sum_{x_0}\Pi_\Gamma^\infty(x-x_0),
\label{eq:FVE_wrap}
\end{equation}
where the sum over $x_0$ runs over
\begin{equation}
x_0\in\{(\pm L,0,0,0), (0,\pm L,0,0), (0,0,\pm L,0), (0,0,0,\pm T), (\pm L,\pm L,0,0),\cdots\}.
\end{equation}

Since the pseudoscalar correlator is expected to contain large finite volume
effects due to the pion pole, we focus on this channel.
The second term on the RHS of (\ref{eq:FVE_wrap}) is then dominated by
the contribution of pion, which is well approximated using the modified Bessel function
$K_1$.
We apply the subtraction
\begin{equation}
\Pi_P^\infty(x)
= \Pi_P^{L^3\times T}(x)
- {z_0M_\pi^2\over2\pi^2}\sum_{x_0}{K_1(M_\pi|x-x_0|)\over|x-x_0|},
\label{eq:FVE_subt}
\end{equation}
where $M_\pi$ and $z_0$ are extracted from the asymptotic form of the
zero-momentum correlator, $\int\td^3x\Pi_P(\vec x,t)\rightarrow z_0\e^{-M_\pi t}$.

\begin{figure}[t]
\begin{center}
\includegraphics[width=120mm]{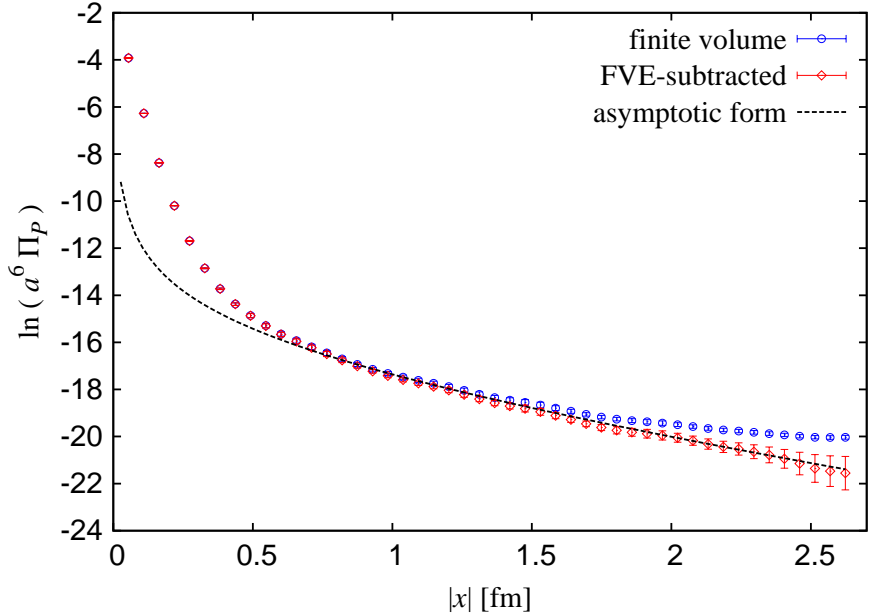}
\caption{
Pseudoscalar correlator
in the time-direction ($|x| = t$) before (circles) and after (diamonds) applying the
subtraction (\ref{eq:FVE_subt}).
Dashed curve corresponds to the asymptotic form
obtained by analyzing the zero momentum correlator.
Data on the same ensemble as in Fig.~\ref{fig:x2vspp_raw} are plotted.
}
\label{fig:dept}
\end{center}
\end{figure}

Figure~\ref{fig:dept} shows the pseudoscalar correlator before
and after applying the
subtraction (\ref{eq:FVE_subt}) of the finite volume effects.
The correlators only on the time axis $(\vec0,t(=|x|))$ are plotted.
We observe that the correlator after applying the subtraction
(\ref{eq:FVE_subt}) becomes consistent with the asymptotic form (dashed curve)
at long distances.

\begin{figure}
\begin{center}
\includegraphics[width=120mm]{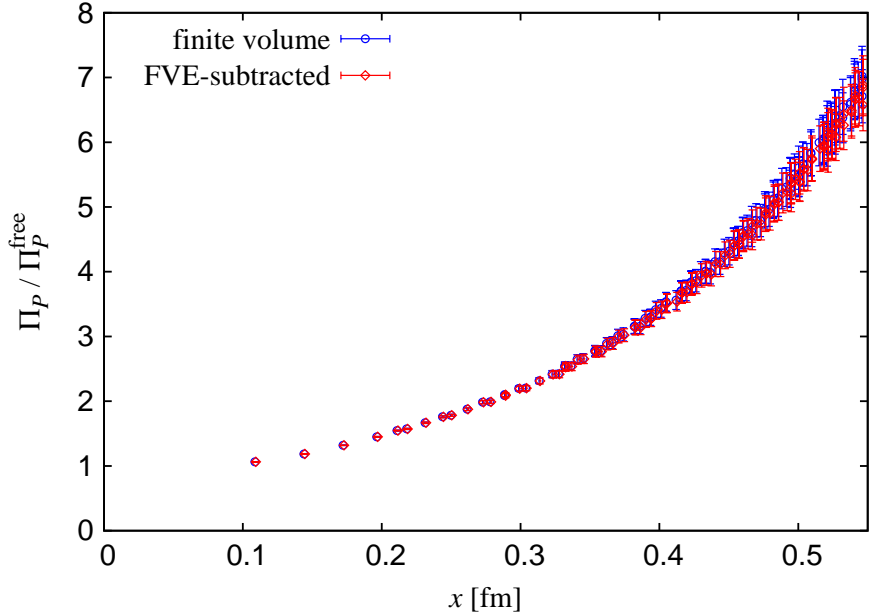}
\caption{
Pseudoscalar correlator before (circles) and
after (diamonds) applying the subtraction (\ref{eq:FVE_subt}).
Data on the same ensemble as in Fig.~\ref{fig:x2vspp_raw} are plotted.
}
\label{fig:fvc_short}
\end{center}
\end{figure}

From Fig.~\ref{fig:dept}, we find that the finite volume effects are
significant at $x\gtrsim1$~fm.
In Fig.~\ref{fig:fvc_short}, we compare $\Pi_P(x)/\Pi_P^{\rm free}(x)$ before
and after the subtraction (\ref{eq:FVE_subt}) in the short distances
$x \lesssim 0.5$~fm and find that
the magnitude of the finite
volume effect in the pseudoscalar correlator at the short distances is less than
20\% of the statistical errors.
Since the finite volume effects are already small in the pseudoscalar correlator,
those in other channels are expected to be negligible.
In this work, we apply the subtraction (\ref{eq:FVE_subt}) only for the
pseudoscalar channel.

\section{Determination of renormalization constants}
\label{sec:NPR}

\subsection{Determination of $Z_V$}
\label{subsec:Zv}

With domain-wall fermions that precisely satisfies the Ginsparg-Wilson
relation, the identity
$Z_V^{\rm\overline{MS}/lat}(a) = Z_A^{\rm\overline{MS}/lat}(a)$ is valid.
In the analysis of $Z_V^{\rm\overline{MS}/lat}(a)$, the renormalization scale 2~GeV
can be omitted because the current conservation ensures its scale independence.

\begin{figure}[t]
\begin{center}
\includegraphics[width=120mm]{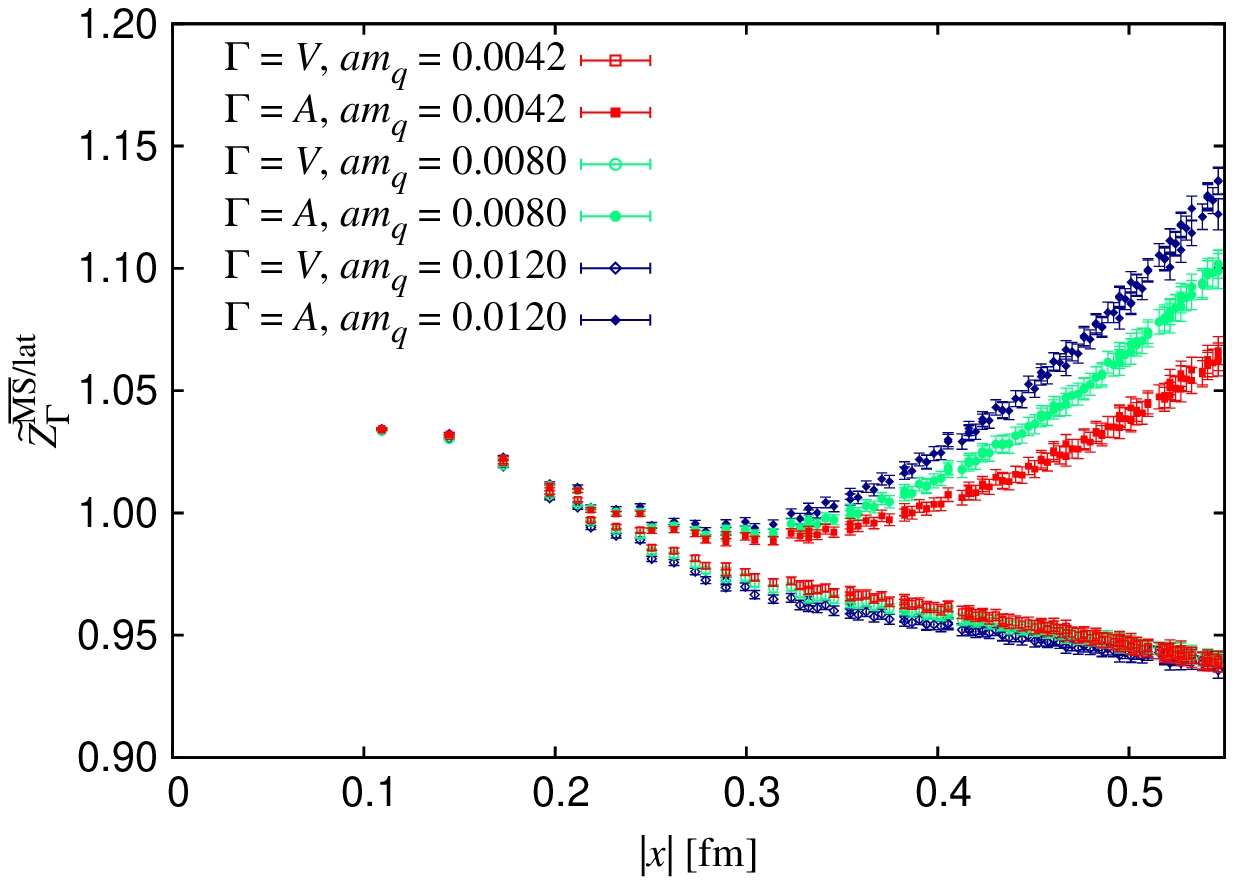}
\caption{
$\widetilde Z^{\rm\overline{MS}/lat}_V(a;x)$ (open points) and
$\widetilde Z_A^{\rm\overline{MS}/lat}(a;x)$ (filled points)
calculated by (\ref{eq:zfactor_cond}) at the three input masses
$am_q = 0.0042$ (squares), 0.080 (circles), 0.0120 (diamonds)
and $\beta = 4.35, am_s = 0.0180$.
}
\label{fig:Zva}
\end{center}
\vspace{4.2mm}
\begin{center}
\includegraphics[width=120mm]{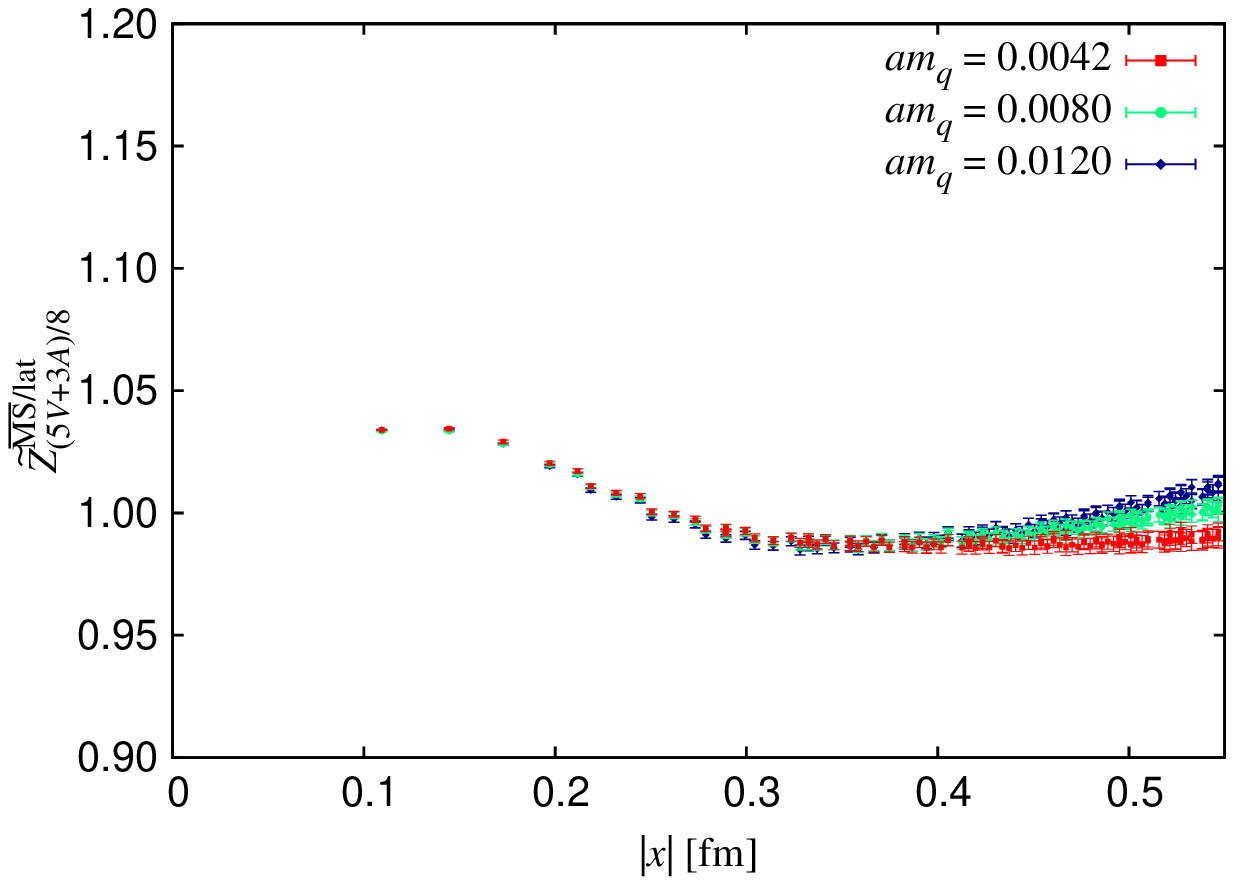}
\caption{
$\widetilde Z_{(5V+3A)/8}^{\rm\overline{MS}/lat}(a;x)$ defined by
(\ref{eq:Z_3v5a}).
The results on the same ensembles as in Fig.~\ref{fig:Zva} are shown.
}
\label{fig:CZv}
\end{center}
\end{figure}

Figure~\ref{fig:Zva} shows $x$-dependence of
$\widetilde Z_V^{\rm\overline{MS}/lat}(a;x)$ and
$\widetilde Z_A^{\rm\overline{MS}/lat}(a;x)$,
which are defined by (\ref{eq:zfactor_cond}), at three input masses
and $\beta = 4.35, am_s = 0.0180$.
For $|x| < 0.2$~fm, the results increase toward the short-distance regime
due to the remnant discretization effects as discussed later.
For $|x| > 0.25$~fm, there is a significant splitting between the vector
and axial-vector channels due to the non-perturbative effects.

The leading non-perturbative effect is described by OPE.
According to the discussion in Section~\ref{subsec:OPEforms},
the coefficients $c_{4,\bar qq}^V$ and $c_{4,\bar qq}^A$ in
the OPE of the vector and axial-vector correlators,
\begin{equation}
\Pi_{V/A}(x)
= {c_0\over x^6}
+ {c_{4,\bar qq}^{V/A}m_q\langle\bar qq\rangle
+ c_{4,G}^{V/A}\langle GG\rangle\over x^2} + \cdots,
\end{equation}
satisfy $c_{4,\bar qq}^V/c_{4,\bar qq}^A=-3/5$ at tree level.
The combination ${1\over8}(5\Pi_V+3\Pi_A)$ is therefore expected to
cancel the leading contribution of the chiral condensate.
Therefore we analyze
\begin{equation}
\widetilde Z_{(5V+3A)/8}^{\rm\overline{MS}/lat}(a;x)
=
\sqrt{\Pi_V^{\rm\overline{MS}}(a;x)
\over
{1\over8}\big(5\Pi_V^{\rm lat}(a;x)+3\Pi_A^{\rm lat}(a;x)\big)},
\label{eq:Z_3v5a}
\end{equation}
to extract the renormalization constant suppressing the non-perturbative effect.
We find that both the $|x|$-dependence and the mass dependence are dramatically
reduced as shown in Fig.~\ref{fig:CZv}.
Although there is no mass dependence remaining for
$\widetilde Z_{(3V+5A)/8}^{\rm\overline{MS}/lat}(a;x)$ in the OPE for the
operators of dimension four, the data still have sizable mass dependence for
$|x| > 0.4$~fm.
This mass dependence may originate from higher
dimensional operators including
$m_q^2\langle GG\rangle$ and $m_q^3\langle\bar qq\rangle$.
Another mass-independent operator $\langle\bar qq\bar qq\rangle$ with
the same mass dimension should also be considered.

\begin{figure}[t]
\begin{center}
\includegraphics[width=120mm]{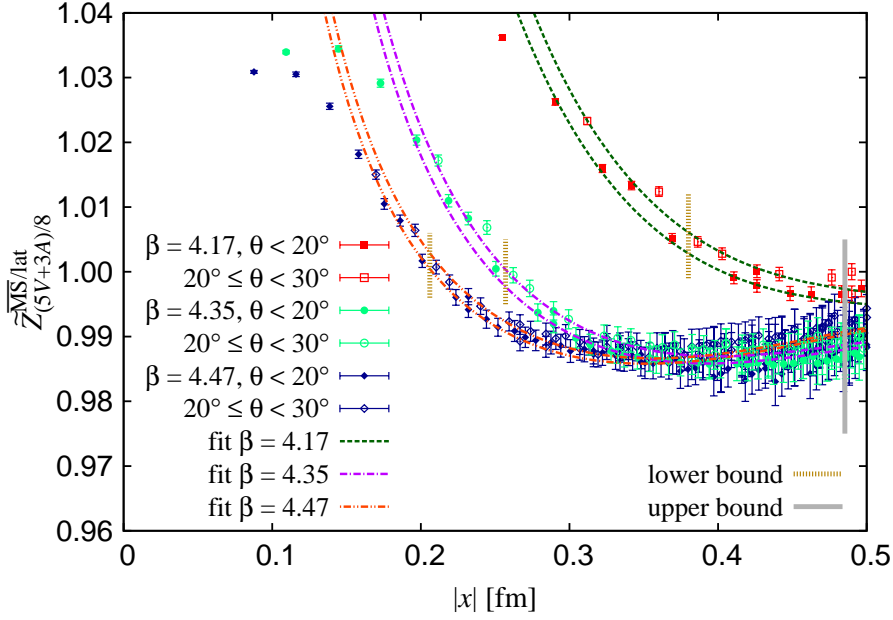}
\caption{
$\widetilde Z_{(5V+3A)/8}^{\rm\overline{MS}/lat}(a;x)$ at the three of $\beta$
with $M_\pi \sim 300$ MeV.
For each $\beta$, the results at smaller $am_s$ is plotted.
The data at 
$\theta < 20^\circ$ (filled points) and
$20^\circ \le \theta < 30^\circ$ (open points)
are separately plotted.
The fit results for both $\theta < 20^\circ$ and $20^\circ \le \theta < 30^\circ$
are also plotted by the curves.
}
\label{fig:CZv_beta}
\end{center}
\end{figure}

Figure~\ref{fig:CZv_beta} shows
$\widetilde Z_{(5V+3A)/8}^{\rm\overline{MS}/lat}(a;x)$ obtained at three
$\beta$ values with approximately matched quark masses.
The position where $\widetilde Z_{(5V+3A)/8}^{\rm\overline{MS}/lat}(a;x)$
starts deviating from a constant toward short distances moves as the lattice
spacing is reduced, indicating that this deviation is due to the discretization effects.
The most significant discretization effect is of $O(a^2)$ which appears as
$(a/x)^2$.
Since we already subtract discretization effects at the tree-level, 
$\alpha_s(\beta)(a/x)^2$ is the leading remaining discretization effects.
As discussed in Section~\ref{subsec:free_correc},
we discard the data in $\theta \ge 30^\circ$ and
parametrize the discretization effects in
$0^\circ \le \theta < 20^\circ$ and in $20^\circ \le \theta < 30^\circ$
separately as described below.

We determine $Z_V^{\rm\overline{MS}/lat}$ by a simultaneous
fit of the data on all ensembles using the fit function
\begin{align}
\widetilde Z_{(5V+3A)/8}^{\rm\overline{MS}/lat}(a;x)
&= Z_V^{\rm\overline{MS}/lat}(\beta)
\notag\\*
&+ C_{-2}(\theta) \alpha_s(\beta) (a/x)^2 + C_{4,G}x^4
+ (C_{6,q} + C_{6,mG} m_q^2 + C_{6,mq}m_q^3)x^6,
\label{eq:fitfunc}
\end{align}
with nine free parameters
$Z_V^{\rm\overline{MS}/lat}(4.17)$,
$Z_V^{\rm\overline{MS}/lat}(4.35)$,
$Z_V^{\rm\overline{MS}/lat}(4.47)$,
$C_{-2}(\theta < 20^\circ)$,
$C_{-2}(20^\circ \le \theta < 30^\circ)$,
$C_{4,G}$ , $C_{6,q}$, $C_{6,mG}$,
and $C_{6,mq}$.
Here, the last four parameters correspond to
the contribution of $\langle a_sGG\rangle$, $\langle\bar qq\bar qq\rangle$,
$m_q^2\langle GG\rangle$, and $m_q^3\langle\bar qq\rangle$, respectively.
In this analysis, we neglect the $O(a_s)$ correction to the Wilson coefficients
of these operators.
The terms of $O(x^6)$ involve the logarithmic dependence
$\ln(x/x_0)^2$ as discussed in Section~\ref{subsec:OPEforms}.
Here we do not consider this effect because $\ln(x/x_0)^2$
is roughly constant in the fit range and their effects cannot be identified.
The fit results are shown in Fig.~\ref{fig:CZv_beta} by the curves.
Here, both the results for $\theta < 20^\circ$ and $20^\circ \le \theta < 30^\circ$
are plotted.

\begin{table}[b]
\caption{
Result for the renormalization factor $Z_V^{\rm\overline{MS}/lat}(a)$ of the vector channel.
}
\label{tab:Zv}
\begin{center}
\begin{tabular}{|c|c|cccc|}
\hline
\multirow{2}{*}{$\beta$} & \multirow{2}{*}{$\ Z_V^{\rm\overline{MS}/lat}(a)$}
& \multicolumn{4}{c|}{Errors} \\ \cline{3-6}
& & Stat. & Disc. & $\mu^*_x$ & $\Lambda_{\rm QCD}$ \\
\hline
\\[-8mm]
\ 4.17\ \ & \ 0.9553 & (53) & (74) & (8) & (5) \\[0mm]
\ 4.35\ \ & \ 0.9636 & (34) & (46) & (7) & (4) \\[0mm]
\ 4.47\ \ & \ 0.9699 & (26) & (38) & (6) & (4) \\[0mm]
\hline
\end{tabular}
\end{center}
\end{table}

In Table~\ref{tab:Zv}, the results for $Z_V^{\rm\overline{MS}/lat}$ are summarized
with the errors from various sources.
The first error is the statistical error.
The second represents the discretization error.
The central value is the fit result with a lower bound $(x_{\rm low}/a)^2 = 23$,
which is shown in Fig.~\ref{fig:CZv_beta}.
and the second error is estimated by moving the lower bound in the region
$19 \le (x_{\rm low}/a)^2 \le 27$ and taking the largest difference from
the central value.
The third is an estimate of the uncertainty of the higher order corrections of the
perturbative expansion as discussed in Section~\ref{subsec:massless_pt}.
The central value is calculated with $\mu_x^{*,\rm opt}$ in (\ref{eq:vv_best_l})
and the uncertainty is estimated by the maximum difference
of results with $\mu^*_x$ in the region
$[{1\over2}\mu_x^{*,\rm opt},2\mu_x^{*,\rm opt}]$.
The last error is from the uncertainty of
$\Lambda_{\rm QCD}$ or of the strong coupling constant.
We use the value
$\Lambda_{\rm QCD}^{{\rm\overline{MS}},n_f=3} = 340(8)$ MeV
reported by Particle Data Group \cite{Agashe:2014kda}.
The uncertainty of the lattice spacing does not significantly affect
the results.
The upper bound of the fit range is fixed to 0.485~fm.

The fit result for other parameters reads
\begin{equation*}
\begin{array}{ll}
C_{-2}(\theta < 20^\circ) = 14.0 (1.2) (2.0) (1) (1),
&C_{-2}(20^\circ \le \theta < 30^\circ) = 15.1 (1.1) (2.2) (1) (\--),
\\
C_{4,G} = 0.564 (190) (187) (99) (14) \rm\ fm^{-4},
&C_{6,q} = -0.109 (61) (56) (4) (3) \rm\ fm^{-6},
\\
C_{6,mG} = -19.8 (31.5) (5) (3) (1) \rm\ fm^{-4},
&C_{6,mq} = 192 (125) (2) (2) (1) \rm\ fm^{-3}.
\end{array}
\end{equation*}
The consistency of this analysis can be checked by evaluating the
gluon condensate from the fit result.
Using (\ref{eq:c_G2}) and the fit result
$C_{4,G}$, we obtain
$\langle(\alpha_s/\pi)GG\rangle = 0.017 (6) (6) (3)$ $(\--) \rm\ GeV^4$
at the lowest order of $\alpha_s/\pi$.
The errors are estimated in the similar manner.
This result is in good agreement with known values, {\it e.g.} 
$\langle(\alpha_s/\pi)GG\rangle = 0.012 \rm\ GeV^4$ from the sum
rule for charmonium \cite{Shifman:1978bx} and
$\langle(\alpha_s/\pi)GG\rangle = 0.006 (12) \rm\ GeV^4$ from the spectral
functions of hadronic $\tau$ decays \cite{Geshkenbein:2001mn}.

\subsection{Determination of $Z_S$}
\label{subsec:Zs}

The domain-wall fermion also guarantees the agreement of the renormalization
constants of the scalar  $Z_S^{\rm\overline{MS}/lat}(a)$ and pseudoscalar
$Z_P^{\rm\overline{MS}/lat}(a)$ densities.
The determination of $Z_S^{\rm\overline{MS}/lat}(a)$ and
$Z_P^{\rm\overline{MS}/lat}(a)$ may be more complicated
due to the instanton-induced 't Hooft interactions
\cite{'tHooft:1976fv,Novikov:1981xi}, which affect the scalar and pseudoscalar
correlators significantly and are not described by OPE.
Since the instanton effects to the scalar and pseudoscalar correlators are the same
magnitude with an opposite sign, the na\"\i ve average
${1\over2}(\Pi_S(x)+\Pi_P(x))$ may cancel such effects and could be
well explained by OPE.
The average contains the contribution of the chiral condensate in OPE,
which we try to cancel by using the difference between the vector and axial-vector
correlators.
Namely, we analyze
\begin{align}
\widetilde Z_{(S+P)/2 + (V-A)/16}^{\rm\overline{MS}/lat}&({\rm2\ GeV},a;x)
\notag\\*
=& \sqrt{\Pi_S^{\rm\overline{MS}}({\rm2\ GeV};x)
\over
{1\over2}\big(\Pi_S^{\rm lat}(a;x) + \Pi_P^{\rm lat}(a;x)\big)
+{1\over16}\big(\Pi_V^{\rm lat}(a;x) - \Pi_A^{\rm lat}(a;x)\big)
},
\label{eq:Z_1s1p}
\end{align}
whose OPE does not depend on the chiral condensate
$m_q\langle\bar qq\rangle x^4$ at the tree-level.
Since we neglect the $O(a_s)$ correction to the Wilson coefficients
in this analysis, we omit the renormalization factor for
$\Pi_V^{\rm lat}(a;x) - \Pi_A^{\rm lat}(a;x)$.

\begin{figure}[t]
\begin{center}
\includegraphics[width=120mm]{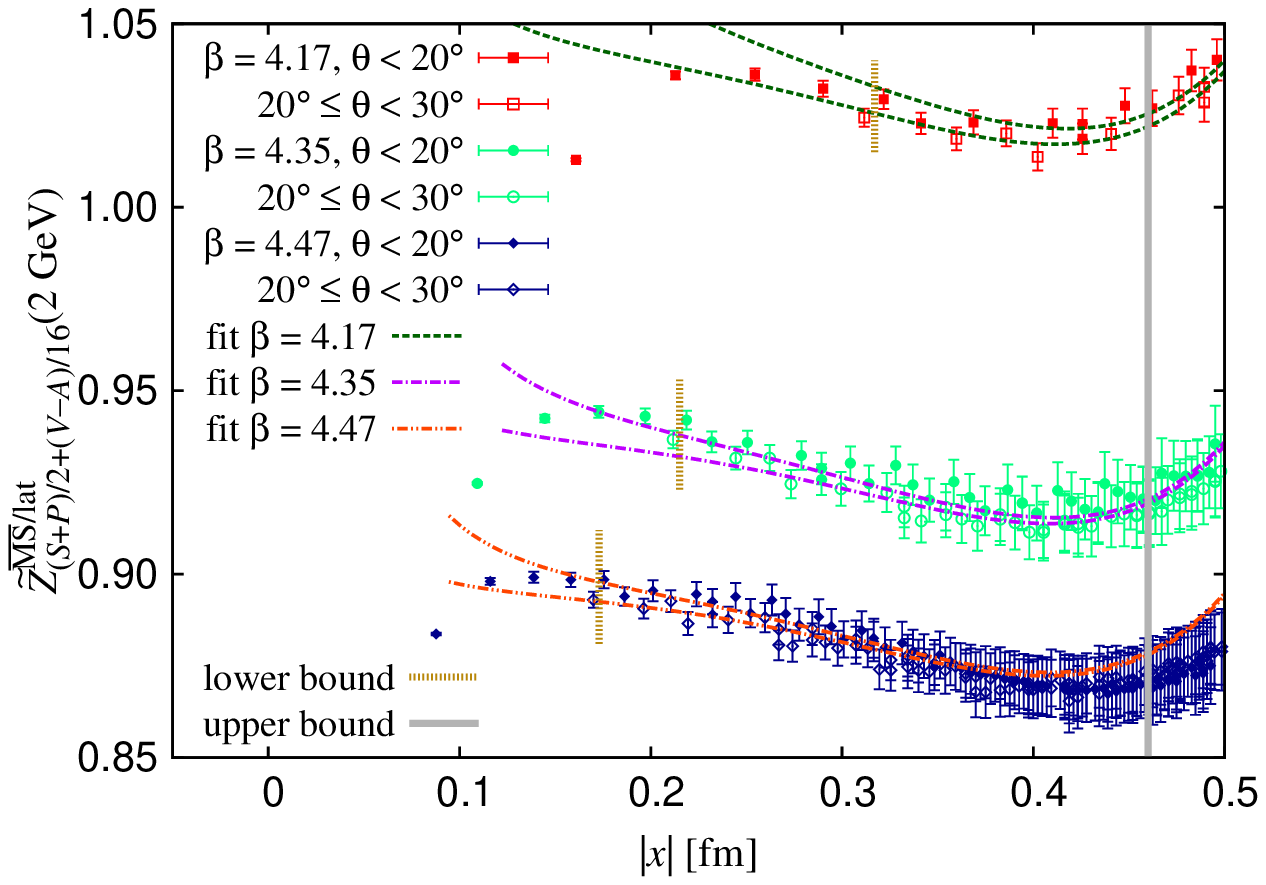}
\caption{
Same as Fig.~\ref{fig:CZv_beta} but for
$\widetilde Z_{(S+P)/2 + (V-A)/16}^{\rm\overline{MS}}({\rm2\ GeV},a;x)$
}
\label{fig:Zs_beta}
\end{center}
\end{figure}

\begin{table}[b]
\caption{
Result for the renormalization factor $Z_S^{\rm\overline{MS}/lat}({2\rm\ GeV},a)$ of the scalor channel.
}
\label{tab:Zs}
\begin{center}
\begin{tabular}{|c|c|cccc|}
\hline
\multirow{2}{*}{$\beta$} & \multirow{2}{*}{$\ Z_S^{\rm\overline{MS}/lat}({2\rm\ GeV},a)$}
& \multicolumn{4}{c|}{Errors} \\ \cline{3-6}
& & Stat. & Disc. & $\mu_x^*$ & $\Lambda_{\rm QCD}$ \\
\hline
\\[-8mm]
\ 4.17\ \  & \ 1.0372 & (93) & (77) & (57) & (58)\\[0mm]
\ 4.35\ \  & \ 0.9342 & (57) & (43) & (37) & (34)\\[0mm]
\ 4.47\ \  & \ 0.8926 & (41) & (35) & (30) & (25)\\[0mm]
\hline
\end{tabular}
\end{center}
\end{table}

We implement the simultaneous fit to the data of (\ref{eq:Z_1s1p}) with the same function
as (\ref{eq:fitfunc}).
The results are summarized in Table~\ref{tab:Zs}.
The error estimation is done in the similar manner as that for the vector channel.
We choose the lower bound of the fit range as $(x_{\rm low}/a)^2 = 16$
for the central value and estimate the second
error by changing $x_{\rm low}$ in the region $12 \le (x_{\rm low}/a)^2 \le 20$.
The central $\mu_x^* = \mu_x^{*,\rm opt}$ is given by (\ref{eq:ss_best_l}) and
the third error is estimated by varying $\mu_x^*$ in the region
$[{1\over1.6} \mu_x^{*,\rm opt}, 1.6 \mu_x^{*,\rm opt}]$.
The upper bound of the fit range is fixed to 0.460~fm.
The fit result is shown in Fig.~\ref{fig:Zs_beta} for the same ensembles as
in Fig.~\ref{fig:CZv_beta}.

The fit result for other parameters reads
\begin{equation*}
\begin{array}{ll}
C_{-2}(\theta < 20^\circ) = 2.25 (1.37) (1.81) (58) (77),
&C_{-2}(20^\circ \le \theta < 30^\circ) = 0.522 (1.39) (1.72) (60) (80),
\\
C_{4,G} = -2.27 (52) (25) (23) (33) \rm\ fm^{-4},
&C_{6,q} = 7.71 (52) (1.08) (2.19) (45) \rm\ fm^{-6},
\\
C_{6,mG} = 212 (156) (4) (3) (2) \rm\ fm^{-4},
&C_{6,mq} = -432 (622) (18) (7) (3) \rm\ fm^{-3}.
\end{array}
\end{equation*}

\section{Conclusion}
\label{sec:summary}

We have determined the renormalization factors $Z_V^{\rm\overline{MS}/lat}(a)$,
$Z_S^{\rm\overline{MS}/lat}({2\rm\ GeV},a)$ of the flavor
non-singlet quark bilinear operators composed of M\"obius domain-wall fermions
by analyzing correlation functions in the coordinate space.
This method enables us to renormalize in a fully gauge invariant manner
and to implement the perturbative matching up to the four-loop level.
We impose the direct renormalization condition onto the $\rm\overline{MS}$
scheme without introducing any intermediate renormalization schemes.

Although the complicated analysis is required to eliminate
discretization effects that are significant at short distances,
the results obtained in this work and their accuracy demonstrate the
use of the X-space method.
It is already at a competing level with the RI/MOM methods, and we expect
it becomes more so in the future as the
lattice spacing is reduced, because the perturbative error would become
the dominant source in such situation.

Direct applications of our results are the calculation of meson decay
constants and quark masses.
Some preliminary results are found in \cite{Fahy:2015xka}.
The (pseudo)scalar renormalization factor is also necessary for the
determination of the chiral condensate, which characterizes the spontaneous
broken chiral symmetry in QCD.
We can extract this important quantity from the same set of
correlators as we analyzed in this work, through the axial Ward-Takahashi identity.
Such analysis will be presented in a forthcoming paper.
The same quantity can be obtained with a completely different method,
{\it i.e.} from the eigenvalue density of the Dirac operator through the
Banks-Casher relation.
A preliminary analysis on our ensembles was presented in
\cite{Cossu:2016yzp}.

Current correlators at short distances contain more information of QCD.
They are complicated mixture of excited states of hadrons, but at
the same time they can be analyzed using perturbation theory.
For the analysis of the intermediate region between perturbative and
non-perturbative regimes, the lattice data may provide useful information,
that can be used to study the range of application of perturbative expansion in
perturbative QCD for instance.
The work in such directions are in progress.

\begin{acknowledgments}
Numerical simulations are performed on Hitachi SR 16000 and IBM System
Blue Gene Solution at KEK under a support of its Large Scale Simulation Program
(No.~13/14-04, 14/15-10).
We thank P. Boyle for the optimized code for BGQ.
This work is supported in part by the Grant-in-Aid of the Japanese Ministry of
Education (No.~25800147, 26247043, 26400259,15K05065) and by MEXT
SPIRE and JICFuS.
\end{acknowledgments}

\appendix

\section{Summary of perturbative coefficients}
\label{sec:app_pert}

This appendix summarizes numerical coefficients of perturbative expansions
of the vector and scalar correlators obtained by using the four-loop results
of correlators \cite{Chetyrkin:2010dx},
the beta function \cite{vanRitbergen:1997va},
and the anomalous dimension \cite{Chetyrkin:1997dh,Vermaseren:1997fq}.

For the vector channel, the perturbative coefficients $C_i^V$'s at
$\mu_x^* = \mu_x$ and the corresponding perturbative expansion with
$n_f = 3$ are
\begin{align}
&C_1^V(\mu_x) = 1,
\notag\\
&C_2^V(\mu_x) = -5.5269 +0.34002n_f,
\notag\\
&C_3^V(\mu_x) = 10.415 -3.2912n_f +0.12602n_f^2,
\notag\\
&C_4^V(\mu_x) = 28.928 +30.966n_f -2.6652n_f^2 +0.064007n_f^3,
\end{align}
\begin{align}
\Pi_V^{\rm\overline{MS}}(x)\Big|_{n_f=3}
={6\over\pi^4x^6}\Big(
1&+a_s(\mu_x) -4.5069a_s(\mu_x)^2
\notag\\
&+1.6758a_s(\mu_x)^3
+99.567a_s(\mu_x)^4
+O(a_s^5)
\Big).
\label{eq:vv_l0.0_nf3}
\end{align}
At $\mu_x^* = \tilde\mu_x$, $C^V_i$'s and the expansion are given by
\begin{align}
&C_1^V(\tilde \mu_x) = 1,
\notag\\
&C_2^V(\tilde \mu_x) = -4.8893 +0.30137n_f,
\notag\\
&C_3^V(\tilde \mu_x) = 5.2517 -2.6633n_f +0.10124n_f^2,
\notag\\
&C_4^V(\tilde \mu_x) = 33.562 +26.233n_f -2.2001n_f^2 +0.050863n_f^3,
\end{align}
\begin{align}
\Pi_V^{\rm\overline{MS}}(x)\Big|_{n_f=3}
= {6\over\pi^4x^6}\Big(
1+&a_s(\tilde\mu_x) -3.9852a_s(\tilde\mu_x)^2
\notag\\
-&1.8270a_s(\tilde\mu_x)^3+93.835a_s(\tilde\mu_x)^4
+O(a_s^5)
\Big).
\label{eq:vv_MStilde_nf3}
\end{align}
Finally, at $\mu_x^* = \tilde\mu_x$, $C^V_i$'s and the expansion are
\begin{align}
&C_1^V(\mu_x^{\rm BLM}) = 1,
\notag\\
&C_2^V(\mu_x^{\rm BLM}) = 0.083333,
\notag\\
&C_3^V(\mu_x^{\rm BLM}) = -7.1191 -1.1478n_f +0.010414n_f^2,
\notag\\
&C_4^V(\mu_x^{\rm BLM}) = -56.886 +12.283n_f -0.58326n_f^2 +0.014075n_f^3,
\end{align}
\begin{align}
\Pi_V^{\rm\overline{MS}}(x)\Big|_{n_f=3}
= {6\over\pi^4x^6}\Big(
1+&a_s(\mu_x^{\rm BLM}) +0.083333a_s(\mu_x^{\rm BLM})^2
\notag\\
-&10.469a_s(\mu_x^{\rm BLM})^3-24.907a_s(\mu_x^{\rm BLM})^4
+O(a_s^5)
\Big).
\label{eq:vv_BLM_nf3}
\end{align}

For the scalar channel, the perturbative coefficients
$C_i^S$'s at $\mu_x^* = \mu_x' = \mu_x$
and the corresponding expansion with $n_f=3$ are
\begin{align}
&C_1^S(\mu_x,\mu_x) = 0.20294,
\notag\\
&C_2^S(\mu_x,\mu_x) = -20.197 +0.56314n_f,
\notag\\
&C_3^S(\mu_x,\mu_x) = 7.8854 -7.5318n_f +0.37635n_f^2,
\notag\\
&C_4^S(\mu_x,\mu_x) = 500.95 +40.402n_f -5.3403n_f^2 +0.18479n_f^3,
\end{align}
\begin{align}
\Pi_S^{\rm\overline{MS}}(\mu_x;x)\Big|_{n_f=3}
= {3\over\pi^4x^6}\Big(
1 &+ 0.20294a_s(\mu_x) -18.507a_s(\mu_x)^2
\notag\\
&-11.323a_s(\mu_x)^3 +579.08a_s(\mu_x)^4
+ O(a_s^5)
\Big).
\label{eq:ss_l0.0_t0.0_nf3}
\end{align}
At $\mu_x^* = \mu_x' = \tilde\mu_x$, $C_i^S$'s and the expansion are given by
\begin{align}
&C_1^S(\tilde\mu_x,\tilde\mu_x) = {2\over3},
\notag\\
&C_2^S(\tilde\mu_x,\tilde\mu_x)
= -17.766 +0.48193n_f,
\notag\\
&C_3^S(\tilde\mu_x,\tilde\mu_x)
= -14.656 -6.3172n_f +0.32333n_f^2,
\notag\\
&C_4^S(\tilde\mu_x,\tilde\mu_x)
= 450.45 +25.502n_f -3.8057n_f^2 +0.14697n_f^3,
\end{align}
\begin{align}
\Pi_S^{\rm\overline{MS}}(\tilde\mu_x;x)\Big|_{n_f=3}
= {3\over\pi^4x^6}\Big(
1 &+ 0.66667a_s(\tilde\mu_x) -16.321 a_s(\tilde\mu_x)^2
\notag\\
&-30.698a_s(\tilde\mu_x)^3 + 496.67a_s(\tilde\mu_x)^4
+ O(a_s^5)
\Big).
\label{eq:pt_ss_tilde_nf3}
\end{align}

\section{Mean field approximation of correlators on the lattice}
\label{sec:free_corr}

Correlators in the coordinate space are given by
\begin{equation}
\Pi_\Gamma(x) = \left\langle\Tr\big[S_F(x)\Gamma S_F(-x)\Gamma\big]\right\rangle,
\end{equation}
where $S_F(x)$ stands for the propagator of the Dirac field.
We give the mean field approximation of the domain-wall propagator
and its asymptotic form in the long-distance limit.
Since the residual mass in this work is almost negligible,
we consider the domain-wall propagator with the infinite size of the fifth direction
$L_s\rightarrow\infty$.
The four-dimensional representation $\widetilde S_F^{\rm DW}(q,m_q)$ of
the mean field domain-wall propagator in the momentum space is
\cite{Narayanan:1992wx,Shamir:1993zy,Aoki:1997xg}
\begin{equation}
\widetilde S_F^{\rm DW}(q,m_q)/a
= {-\img\gamma_\mu u_0\sin (aq_\mu) + (1-W\e^{-\alpha})am_q
\over -1 + W\e^\alpha + (1-W\e^{-\alpha})(am_q)^2},
\label{eq:DWqprop}
\end{equation}
where $W$ and $\alpha$ are defined by
\begin{align}
W(q) &= 1 - M_0 + \sum_\mu\big(1-u_0\cos(aq_\mu)\big),
\\
\cosh\alpha(q) &= {1+ W^2 + u_0^2\sum_\mu\sin^2(aq_\mu)\over 2W},
\end{align}
with $M_0$ and $u_0$ being the domain-wall mass parameter and
the fourth root of the plaquette expectation value, respectively.
The free propagator is reproduced by putting $u_0=1$.

The propagator in the coordinate space is calculated by a numerical
Fourier transform in a finite box $L^4$,
\begin{equation}
S_F^{{\rm DW},L^4}(x,m_q)
= {1\over L^4}\sum_{q} \widetilde S_F^{\rm DW}(q) \e^{\img qx},
\end{equation}
where the sum over $q$ for the periodic boundary condition runs over
\begin{equation}
q \in
\left\{
{2\pi\over L}(k_1,k_2,k_3,k_4)\ \Big|\ 
k_\mu = -{L\over2a}+1,-{L\over2a}+2,\cdots, {L\over2a}
\right\},
\end{equation}

The mean field approximation of the propagator of a light quark in a finite box
$(m_qL \lesssim 1)$ involves large finite volume effects because quarks
are in the deconfining phase.
Since the numerical Fourier transform in sufficient large volumes is expensive,
we apply a correction for the domain-wall propagator
as we applied for the propagators of bosonic fields in Section~\ref{subsec:FVE},
\begin{align}
S_F^{\rm DW,\infty}(x,m_q)
&= S_F^{{\rm DW}, L^4}(x,m_q) - \sum_{x_0}S_F^{\rm DW,\infty}(x-x_0,m_q)
\notag\\
&\simeq S_F^{{\rm DW}, L^4}(x,m_q) - \sum_{x_0}S_F^{\rm DW,asym}(x-x_0,m_q),
\label{eq:subt_DiracFVE}
\end{align}
where the sum over $x_0$ runs over
\begin{equation}
x_0 \in \{(\pm L,0,0,0), (0,\pm L,0,0), (0,0,\pm L,0), (0,0,0,\pm L), (\pm L,\pm L,0,0),\cdots\},
\end{equation}
and $S_F^{\rm DW,asym}(x)$ is the asymptotic form of the domain-wall propagator
in the long-distance limit in the infinite volume, which is calculated as follows.

The domain-wall propagator in the low-momentum region
is calculated by expanding (\ref{eq:DWqprop}) at $aq \simeq 0$,
\begin{equation}
\widetilde S_F^{\rm DW}(q,m_q) \xrightarrow{aq\rightarrow0}
(1-\delta^2){-\img u_0\slashchar{q} + m'_q\over u_0^2q^2+{m'_q}^2}
+ O(\delta^4),
\end{equation}
with
\begin{align}
\delta &= 1 - M_0 + 4(1-u_0),
\\
m'_q &= (1-\delta^2)m_q.
\end{align}
The Fourier transform of this propagator in the infinite volume gives
an asymptotic form in the long-distance limit of the domain-wall propagator,
\begin{equation}
S_F^{\rm DW,asym}(x,m_q)
= {1-\delta^2\over u_0^4}S_F^{\rm cont}(x/u_0,m'_q) + O(\delta^4),
\end{equation}
where the Feynman propagator $S_F^{\rm cont}(x,m_q)$ of a Dirac field in
the continuum coordinate space is given \cite{Zhang:2008jy} by
\begin{align}
S_F^{\rm cont}(x,m_q)
&= \int{\td^4q\over(2\pi)^4}\e^{\img qx}
{-\img \slashchar{q} + m_q\over q^2 + m_q^2}
\notag\\
&= {m_q\slashchar{x}\over4\pi^2|x|^3}K_1(m_q|x|)
+ {m_q^2\slashchar{x}\over8\pi^2x^2}
\big[K_0(m_q|x|) + K_2(m_q|x|)\big]
+{m_q^2\over4\pi^2|x|}K_1(m_q|x|),
\end{align}
with $K_i$ being the modified Bessel functions.

The subtraction by the second line of (\ref{eq:subt_DiracFVE}) cannot
eliminate discretization effects of wrapping effects, {\it i.e.} the
subtraction leaves the discretization effects
\begin{equation}
\sum_{x_0} \big(S_F^{{\rm DW},\infty}(x-x_0,m_q)-S_F^{\rm DW,asym}(x-x_0,m_q)\big),
\end{equation}
which are suppressed in large volumes.
We observe that these discretization effects are sufficiently small
when the calculation is done with $m_q L \gtrsim 1$.

\bibliography{NPR}
\end{document}